\documentclass[twocolumn,english,rmp,showpacs,amsmath,amssymb,eqsecnum]{revtex4}
\usepackage[T1]{fontenc}
\usepackage[latin9]{inputenc}
\usepackage{amsmath}
\usepackage{amssymb}
\usepackage{graphicx}
\usepackage{esint}

\makeatletter
\@ifundefined{textcolor}{}
{%
 \definecolor{BLACK}{gray}{0}
 \definecolor{WHITE}{gray}{1}
 \definecolor{RED}{rgb}{1,0,0}
 \definecolor{GREEN}{rgb}{0,1,0}
 \definecolor{BLUE}{rgb}{0,0,1}
 \definecolor{CYAN}{cmyk}{1,0,0,0}
 \definecolor{MAGENTA}{cmyk}{0,1,0,0}
 \definecolor{YELLOW}{cmyk}{0,0,1,0}
}

\def\be{\begin{equation}}
\def\ee{\end{equation}}
\def\bea{\begin{eqnarray}}
\def\eea{\end{eqnarray}}
\def\bml{\begin{mathletters}}
\def\eml{\end{mathletters}}
\def\bse{\begin{subequations}}
\def\ese{\end{subequations}}
\usepackage{dcolumn}
\usepackage{bm}
\@ifundefined{definecolor}{\usepackage{color}}{}

\usepackage{babel}

\usepackage{babel}

\usepackage{babel}

\usepackage{babel}

\usepackage{babel}

\usepackage{babel}

\makeatother

\usepackage{babel}
\begin{document}

\title{Random First Order Theory concepts in Biology and Condensed Matter physics}

\author{T.R.Kirkpatrick}

\thanks{Electronic address: tedkirkp@umd.edu}

\affiliation{Institute for Physical Science and Technology, and Department of
Physics, University of \\
 Maryland, College Park, Maryland 20742, USA}

\author{D.Thirumalai}

\thanks{Electronic address: dave.thirumalai@gmail.com}

\affiliation{Institute for Physical Science and Technology, and Department of
Chemistry, University of Maryland, College,\\
 Park, Maryland 0742, USA}

\date{\today}
\begin{abstract}
\textit{The routine transformation of a liquid, as it is cooled rapidly,
resulting in glass formation, is remarkably complex. A theoretical
explanation of the dynamics associated with this process has remained
one of the major unsolved problems in condensed matter physics. The
Random First Order Transition (RFOT) theory, which was proposed over
twenty five years ago, provides a theoretical basis for explaining
much of the phenomena associated with glass forming materials. It
links or relates multiple metastable states, slow or glassy dynamics,
dynamic heterogeneity, and both a dynamical and an ideal glass transition.
Remarkably, the major concepts in the RFOT theory can also be profitably
used to understand many spectacular phenomena in biology and condensed
matter physics, as we illustrate here. The presence of a large number
of metastable states and the dynamics in such complex landscapes in biological
systems from molecular to cellular scale and beyond leads to behavior,
which is amenable to descriptions based on the RFOT theory. Somewhat
surprisingly even intratumor heterogeneity arising from variations
in cancer metastasis in different cells is hauntingly similar to glassy
systems. There are also deep connections between glass physics and
electronically disordered systems undergoing a metal-insulator transition,
aging effects in which quantum effects play a role, and the physics
of super glasses (a phase that is simultaneously a super fluid and
a frozen amorphous structure). We argue that the common aspect in
all these diverse phenomena is that multiple symmetry unrelated states
governing both the equilibrium and dynamical behavior - a
lynchpin in the RFOT theory - controls the behavior observed in these
unrelated systems.} 
\end{abstract}
\maketitle
\tableofcontents{}

\section{Introduction}

\label{sec:I} Glasses, which were created over five thousand years
ago in Mesopotamia as objects of astounding beauty, are now used in
everyday life. The unusual properties of this amorphous state makes
glasses ideal materials for use in myriad ways ranging from displays
to architectural solutions to electronics. These applications have
literally altered our lifestyles without us fully appreciating their
utilities or even being aware of them. Despite their ubiquitous presence the fundamental physics
of the process driving their formation has remained elusive \cite{Berthier_Biroli_2011,Parisi_Zamponi_2010}.
The quest to understand the dynamics of the liquid to glass transition
has lead to a number of conceptual ideas, which have been used to
explain a variety of experimental observations. The unabated efforts
to produce a framework to describe the nature of the structural glass
transition (SGT) problem have been summarized in a number of reviews
in the last twenty years \cite{Kirkpatrick_Thirumalai_1995b,Lubchenko07ARPC,Berthier_Biroli_2011,Parisi_Zamponi_2010}.
In this \textit{colloquium}, we first summarize the essential ideas
underlying the random first order transition theory (RFOT) of the
glass transition \cite{Kirkpatrick87PRL,Kirkpatrick87PRB,Kirkpatrick_Thirumalai_Wolynes_1989,Kirkpatrick_Wolynes_1987a,Kirkpatrick_Wolynes_1987b}.
The RFOT theory, a phrase that was first introduced in \cite{Kirkpatrick_Thirumalai_Wolynes_1989},
has previously been used to understand both the structural glass transition
\cite{Kirkpatrick_Thirumalai_1995b} as well as various types of spin
glass transitions without inversion symmetry \cite{Kirkpatrick87PRL,Kirkpatrick87PRB,Kirkpatrick_Wolynes_1987b,Kirkpatrick_Thirumalai_1988a}.
The overarching goal of this work is not to review the current status
of theories for glass physics but is to illustrate how the ideas that
underlie the RFOT can be used to discuss glassy aspects or features
that are manifested in both biological and condensed matter systems.
These applications are meant to show case the wide ranging use of
ideas that were generated in the physics of the SGT. For lack of space
we do not discuss potential connections between the RFOT theory and
other interesting subjects such as turbulence \cite{Dauchot_Bertin_2012,Dauchot_Bertin_2013}.

\subsection{General remarks on glassy systems}

The phenomenology of glasses is well documented. Most liquids, when
undercooled rapidly (0.1 - 100 K/min in the laboratory), become extraordinarily
viscous (Fig. \ref{GlassesGeneralFig1}a) over a narrow temperature
range. The laboratory glass temperature, $T_{g}$, is experimentally
defined when the shear viscosity $\eta(T_{g})\approx10^{13}$ Poise
\cite{Berthier_Biroli_2011}, which is about fifteen orders of magnitude
larger than the viscosity of pure water at room temperature. At $T_g$ the relaxation time becomes so large that thermal equilibrium is not reached in cooling experiments. The dramatic
increase in $\eta$ is not accompanied by discernible changes in the
structure. The temperature-dependent relaxation time scale, $\tau_{\alpha}$
($\approx1/\eta$), is typically super-Arrhenius, and can be fit using
the Vogel-Fulcher-Tamman (VFT) equation, 
\begin{equation}
\tau_{\alpha}=\tau_{0}exp[\frac{D}{(T/T_{0}-1)}]
\label{VFT}
\end{equation}
 where $\tau_{0}$ is a microscopic relaxation time, the parameter
$D$ is referred to as the fragility index, and $T_{0}$ is a putative
ideal glass transition temperature that is obtained by extrapolating measured viscosity data to inaccessible temperatures.  Typically $\alpha$-relaxation, a terminology borrowed from the literature in polymer glasses,  refers to motion on length scales larger than the molecular size of the particles. An example of such a fit for salol
is given in Fig.\ref{GlassesGeneralFig1}c. The viscosity data have
been fit using other forms \cite{Bassler87PRL,Zwanzig88PNAS,Biroli13JCP}
but here we will assume that Eq.(\ref{VFT}) provides a good description,
which indeed is the case for a number of glass forming materials. As shown schematically in Fig. \ref{GlassesGeneralFig1}b,
for many systems the intermediate scattering function exhibits a plateau
after the initial decay before decaying further on the $\alpha$-relaxation
time. The duration of the plateau in regime B (Fig. \ref{GlassesGeneralFig1}b)
increases as the degree of supercooling increases. In addition, the
glassy phase is dynamically heterogeneous - a notion that has received
considerable attention, and here we argue that this concept, arising
naturally from the RFOT, is of great generality. We do not delve into
other interesting aberrations in supercooled liquids, such as the
break down of Stokes-Einstein relation noted in computer simulations
\cite{Barrat90ChemPhys,Thirumalai_Mountain_1993,Shi13JCP} and in
experiments because they are not relevant to the main themes of this
colloquium.

\subsection{Configurational Entropy and the Kauzmann Paradox}

Configurational entropy is a key concept in many theories of the glass
transition. To define it, consider a free energy functional $F(n,T)$
which depends on number density, $n(\mathbf{x})$, and a temperature,
$T$. We assume that at sufficiently low temperatures $F(n,T)$ 
has many minima (that is, the number of minima goes to infinity with
the system volume, $V$). These states are labelled by an index $\gamma$
such that to each valley we associate a free energy $F_{\gamma}$
and a free energy density $f_{\gamma}=F_{\gamma}/V$. The number of
free energy minima with free energy density $f$ is assumed to be
exponentially large:
\begin{equation}
\mathcal{N}(f,T,V)\sim\exp[VS_{c}(f,T)]
\end{equation}
where the function $S_{c}$%
\footnote{More precisely this is the definition of the state entropy or complexity. These quantities can be exactly defined  in mean-field and infinite dimensional models. In more realistic systems we identify the complexity with the configurational entropy.%
}
 is the configurational entropy
or complexity. Physically, it is the entropy arising from an assumed
exponentially large number of locally stable configurations.

The concept of configurational entropy also plays a central role
in experimental glassy physics, especially in the formulation and
analysis of the so-called Kauzmann paradox \cite{Kauzmann48ChemRev}, which follows from the following arguments. Below the melting temperature, $T_{m}$, the heat capacity $C_{p}(T)$
of a supercooled liquid is larger than that of the corresponding crystal.
As a result of this excess heat capacity, the entropy of the supercooled
liquid state is larger than that of the crystal. However, the supercooled
liquid state entropy is decreasing faster than the crystal entropy.
These observations are illustrated in Fig. \ref{Kauzmann} and leads to the 
paradox.  If the entropy difference is extrapolated to temperatures
below the laboratory glass temperature, $T_{g}$, it 
vanishes at some nonzero temperature called $T_{K}$, after
Kauzmann. In his original article Kauzmann \cite{Kauzmann48ChemRev} suggested that the improbable
decrease of the entropy of the liquid below that of the crystal phase could be avoided
if the barrier to nucleation vanished somewhere between $T_{g}$ and
$T_{K}$. Extensive analysis \cite{Angell_1986} shows that the crystallization rate slows
down more rapidly than the relaxation rate so that Kauzmann's paradox
is not realized. One alternative to Kauzmann's suggestion is that
a very slowly cooled liquid would continue to lose entropy until, as it
approaches the crystal value, an equilibrium phase transition occurs.
The nature and the very existence of such a transition has been much debated. Importantly, the extrapolated $T_{K}$ is always close to the fitted $T_{0}$ in Eq.(\ref{VFT}) \cite{Berthier_Biroli_2011}.

The Adams-Gibbs (AG) \cite{Adam_Gibbs_1965} theory of the glass
transition as well as the RFOT theory (below $T_d$ indicated in Fig. \ref{Kauzmann}) focus on the configurational
entropy as defined above. The idea is as follows. Physically, it is reasonable to assume that
that the vibrational part of any amorphous state entropy
should be more or less equal to the crystal state vibrational entropy.
The excess entropy of the liquid state is then attributed to the configurational
entropy, $S_{c}$. The equilibrium phase transition would occur at $T_K$ when
$S_{c}$ vanishes.  In this picture, we expect $T_{K}$ to equal $T_{0}$ because slow transport below $T_d$  is intimately related to loss in $S_{c}$. These ideas will be developed in detail below.

Finally, we remark that both AG and RFOT theories lead to the VFT law, Eq.(\ref{VFT}), via the AG relation,
\begin{equation}
\tau_{\alpha}=\tau_{0}exp[\frac{d}{S_{c}}]\label{AG}
\end{equation}
with $S_{c}\sim\epsilon={\frac{T}{T_K}-1}$, vanishing at $T_{K}$ which is identified as $T_{0}$, and $d$ a positive constant. However, the derivations of Eq.(\ref{AG}) in the two theories are very different. In the AG theory it is concluded  that there is a divergent length scale,  $\xi_{AG}\sim1/|\epsilon|^{1/d}$ while in RFOT the length scale is $\xi_{RFOT}\sim1/|\epsilon|^{2/d}$. $\xi_{RFOT}$ is derived and discussed further below. The AG correlation length exponent $\nu_{AG}=1/d$ is in general not consistent with numerous simulations \cite{Biroli_et_al_2008, Karmakar_et_al_2009, Berthier_Kob_2012, Biroli_et_al_2013} to measure correlations in glassy liquids, nor is it consistent with the expected inequality $\nu\geq2/d$ \cite{Kirkpatrick_Thirumalai_2014}.

\subsection{On the origins of the RFOT}

In an attempt to provide a theory to account for the nature of the
SGT, a framework was developed in the late 1980s, which was initially
inspired%
\footnote{The development was also inspired by an early paper \cite{Kirkpatrick_Wolynes_1987a}
that indicated the MCT of the glass transition is related to a static
density functional description of the glassy state.%
} by analogies to exotic (explained further below) spin glass models
i.e, those without inversion symmetry \cite{Kirkpatrick87PRL,Kirkpatrick87PRB,Kirkpatrick_Wolynes_1987b,Kirkpatrick_Thirumalai_1995a}.
From general physical considerations it is logical that there ought
to be similarities between structural glasses and spin glasses. A
glass, after all, can be thought as a frozen liquid or more precisely
flows on time scales that vastly exceed observational time scale (Fig. \ref{GlassesGeneralFig1}). In both cases there is no obvious long range order. It is only when the
systems evolve in time that there are obvious differences. In a liquid
a particle can diffuse arbitrarily far away from the initial position
as $t\rightarrow\infty$, whereas it would be localized in a small
region in space in a glass on the observation time scale, $\tau_{obs}$
(Fig. \ref{GlassesGeneralFig1}(d)). Similarly, a spin glass may be thought of as a frozen paramagnet
with no long range the magnetic order \cite{Mezard_1987}. When the system
develops in time the local magnetic moment points in a specific average
direction (a spin at a given time remains correlated with itself at
a later time) in the spin glass phase, whereas in the paramagnetic
the spin direction averages to zero resulting in the vanishing of
local magnetic moment. In both cases, it is only through time evolution
can the two phases be distinguished, a concept that will play an important
role in our discussion of dynamic heterogeneity.

Despite the compelling analogy between the structural and spin glasses
there are also important conceptual differences between the two. First,
in SG disorder is quenched \cite{Edwards_Anderson_1975}. In Ising
spin glasses the magnetic moments of Mn in an alloy with Ni are permanently
frozen. On the other hand, in the SGT problem the randomness is self-generated
\cite{Kirkpatrick_Thirumalai_1989} as the material is cooled below
the melting temperature. Second, there is considerable numerical and experimental evidence for an equilibrium phase transition
in three dimensions in Ising spin glasses \cite{Binder86RMP}. In
the SGT case a thermodynamic transition at $T=T_{K}\approx T_{0}$,
characterized by a vanishing of configurational entropy, is not universally
accepted despite considerable experimental and theoretical support.
It is worth emphasizing that the Ising spin glass does not exhibit
unusual slowing down in the relaxation times as a liquid that undergoes
a transition to a supercooled state. Therefore, an analogy to spin
glasses is insightful only if inversion symmetry is not satisfied
as is the case in $p$-spin glass models with $p>2$ \cite{Kirkpatrick87PRL,Gross_Mezard_1984}. 
Indeed, for such models, we that the mathematical
structure of the dynamical equation describing the relaxation of spin-spin
correlation for $p=3$-spin glass model is identical \cite{Kirkpatrick87PRL}
to the Mode Coupling Theory of the density-density relaxation \cite{Goetze_2009,Bengtzelius_et_al_1984,Leutheusser_1984}.
This discovery and subsequent studies linking dynamics and thermodynamics
in these exotic mean-field spin glass models to models in which randomness
is self-generated lead to the complete formulation of the RFOT \cite{Kirkpatrick_Thirumalai_Wolynes_1989}. It is worth noting that the Random Energy Model \cite{Derrida81PRB}, which does not have a dynamical transition at high temperature, is a special case of p-spin glass model with $p = \infty$ exhibiting one step replica symmetry breaking \cite{Gross_Mezard_1984}. 

A valid criticism in using exotic spin glass models to obtain insights
into the SGT is that in the former quenched disorder is explicitly
modeled in the Hamiltonian whereas in the SGT it is self-generated
and manifests itself in the glassy phase. (Because glasses are formed
from molecules, whose dynamics (assuming quantum effects are not important)
at the microscopic level is Newtonian or Brownian, the information
about the self-generated randomness is implicit in the trajectories.)
In what we consider, especially in retrospective, to be an important
paper \cite{Kirkpatrick_Thirumalai_1989}, it was shown that the major
conclusions drawn based on spin-glass models can be obtained using
a model Hamiltonian with short range local order, incorporated using
static structure factor, without any need for explicitly modeling
quenched disorder. We clarified that the generic ideas within RFOT and related theories that produce
a profound connection between static and dynamic description, solely
depends on the emergence of an exponential number of metastable states
at a dynamical transition temperature, $T_{d}$, which is well above
$T_{g}$, the laboratory glass transition temperature. Moreover, the
theory further established that at the so-called temperature Kauzmann
temperature, $T_{K}$, the entropy associated with the metastable
state vanishes resulting in divergence of viscosity Eq.(\ref{VFT}).
Thus, within RFOT $T_{K}=T_{0}$, an observation that accords well
with many glass forming materials \cite{Berthier_Biroli_2011}. 

The body of works created in the late 1980s was subsequently put on
firmer foundations by others \cite{Mezard_Parisi_1996,Parisi_Zamponi_2010,Berthier_Biroli_2011}.
There has also been much discussion  about the sense in which the
mode coupling theory is a proper mean field theory \cite{Andreanov_et_al_2009,Schmid_Schilling_2010,Ikeda_Miyazaki_2010,Franz_et_al_2012}
Very recently \cite{Kurchan_et_al_2012,Kurchan_et_al_2013,Chabonneau_et_al_2013}
all aspects of the RFOT have been illustrated in an exact description
of a hard sphere fluid in the limit of high dimensions. We believe these are illuminating studies because they provide a microscopic basis for understanding RFOT theory of the transition of the SGT, which cannot be unequivocally stated for other theories of supercooled liquids and glasses.

\subsection{Assessment of the RFOT}
 
Despite the successes of the RFOT theory of the SGT, some worrying aspects have been raised  \cite{Biroli_Bouchaud_2012}, which apparently require further scrutiny. Before addressing these issues briefly, it is worth reminding the readers that although the RFOT theory provides a unified description of both the dynamical transition and the expected thermodynamic transition precipitated by the vanishing of the configurational entropy \cite{Kirkpatrick87PRL,Kirkpatrick87PRB,Kirkpatrick_Thirumalai_Wolynes_1989,Kirkpatrick_Thirumalai_1989} much of the focus has been on the dynamics below $T_d$. This is a pity because RFOT theory seamlessly integrates the dynamics above and below $T_d$.  With this in mind we address the difficulties  raised in \cite{Biroli_Bouchaud_2012} associated with RFOT. The first is related to the notion of surface tension between two mosaic states, a concept that is relevant below $T_d$.  As we show below, the crucial element in producing the VFT equation with the vanishing of configurational entropy at $T_K$ is that the free energy barriers between mosaic states scale as $\xi^{d/2}$ where $\xi$  diverges at $T_K$ (see below).  This implies that there is really no interface between two statistical similar glassy states. Rather, to go from one state to another, roughly $N^{1/2}$ of the
$N$ particles in a correlated volume must be rearranged. Consequently, it is our opinion that the issues raised about the surface tension within RFOT merely obfuscates the physics of the activated transitions below $T_d$. The second point is based on the idea that the free energy barriers might depend on the infinite frequency shear modulus, $G_{\infty}$ \cite{Dyre98JNon-Cryst}. Such a postulate has no microscopic basis, which relies on the dubious relation that $\eta = G_{\infty} \tau_{\alpha}$ \cite{Shi13JCP}. Therefore, we feel that the Dyre model cannot be compared with RFOT on the same footing. Finally, it has been asserted in \cite{Biroli_Bouchaud_2012} that the crossover between the liquid like diffusion at $T > T_d$ and the activated transitions below $T_d$ is not only poorly understood but also is without experimental support. This is most certainly not the case as detailed elsewhere \cite{Goetze_2009}. Indeed, unlike the existence of the ideal glass transition temperature, there is much data establishing a change in the dynamics at a temperature far above $T_g$. Moreover, early computer simulations have unequivocally established a change in the nature of transport near $T_d$ \cite{Barrat90ChemPhys,Thirumalai_Mountain_1993,Mountain_Thirumalai_1987}. More importantly, using a detailed analysis of experimental data it has been shown \cite{Novikov03PRE} that not only is there ample evidence for the crossover temperature but also the dynamics leading to the crossover may be semi-universal (see below). We conclude that both experiments and simulations have provided compelling evidence for the existence and importance of $T_d$. In our view the crucial missing point in full support of RFOT is the clear demonstration of the divergent  correlation length $\xi_{RFOT}$ at $T_K$, which was discussed briefly in Section I.B and in more detail below.

\section{Basic notions of the RFOT for the structural glass transition}

In this section, we briefly review the basic features of the RFOT
as applied to the structural glass transition problem.

\subsection{Two transitions}

\label{subsec:II.A-1} The important physical aspects of the glass
transition and the glassy state that are incapsulated in RFOT \cite{Kirkpatrick_Thirumalai_Wolynes_1989,Kirkpatrick_Thirumalai_1989}
are based on two key ideas. First, the glassy state is essentially
a frozen liquid with elastic properties. To describe a glassy state
we imagine an order parameter (OP) description in terms of frozen
density fluctuations, $\delta n=n-n_{l}$. Here $n$ is the statistical
mechanical average local number density and $n_{l}$ is the spatially
averaged density which is identical to the liquid state density at
the same temperature and pressure. In what follows, we will take into
account that there can be many glassy states so that $n$ will have
a state label, $n_{s}$. Other order parameters can be imagined, but
frozen density fluctuations are the simplest and are  directly
related to the most obvious characteristic of a solid: Elastic properties
and a non-zero Debye-Waller factor. Because the glassy phase is amorphous
or has random characteristics, the frozen density OP is specified
by a functional probability measure $DP[\delta n]$ \cite{Kirkpatrick_Thirumalai_1989}.
The first two moments of this measure are, 
\begin{equation}
\overline{\delta n(\mathbf{x})}=\int DP[\delta n]\delta n(\mathrm{\mathbf{x})}\label{eq:average density fluctuation}
\end{equation}
 
\begin{equation}
q\equiv\overline{[\delta n(\mathbf{x})]^{2}}=\int DP[\delta n][\delta n(\mathbf{x})]^{2}.\label{eq:order parameter}
\end{equation}
 Note that because $\overline{n(\mathbf{x})}=n_{l}$ the density itself
cannot be a proper order parameter for the glass transition. In order
to capture the essence of the glassy state, something analogous to  $q$ must
be used \cite{Kirkpatrick_Thirumalai_Wolynes_1989,Kirkpatrick_Thirumalai_1989}.

At the SGT transition the OP will be discontinuous. There are two
arguments leading to this conclusion. First, any Landau or LGW type
theory for the density $n$ will not have a $n\rightarrow-n$ symmetry,
and therefore a Landau type theory will lead to some sort of discontinuous
transition. More formally, the glassy state is one with a broken translational
symmetry, (since it is randomly nonuniform) ,with elastic properties \cite{Palmer82AdvPhys, Szamel11PRL}.  Because of this broken translational symmetry it is impossible to go continuously from a
liquid state with time average translational invariance to a glassy
state. The second argument imagines that for a given glassy state,
the frozen density can be written as \cite{Singh_et_al_1985,Dasgupta99PRE,Ramakrishnan_Yussouff_1979},
\begin{equation}
n(\mathbf{x})\sim\sum_{i}\exp[-(\mathbf{x}-\mathbf{R_{i})^{2}/\mathrm{2<(}}\Delta\mathbf{R)^{2}>]}
\end{equation}
 where the $\{\mathbf{R}_{i}\}$ are the (random or amorphous) average
positions of the particles making up the glassy state and $<(\Delta\mathbf{R})^{2}>$
is the average fluctuation of their positions. In the glassy state
this is a finite quantity, on the order of a particle diameter, and
it determines the Debye-Waller factor and the elastic coefficients
of the glass. In a liquid it grows with time and is proportional to
the self-diffusion coefficient (Fig. 1d). As the glassy state is approached
from the liquid side, it initially grows, but then plateaus on the
scale of a molecular diameter. The plateauing of this mean-squared-displacement
means that the broken translational symmetry of the glassy state will
occur discontinuously. As mentioned earlier, the use of density as
an order parameter \cite{Singh_et_al_1985} differs conceptually from
the ideas of RFOT theory. In the description of the amorphous state
using the density functional theory of the liquid to crystal transition
\cite{Ramakrishnan_Yussouff_1979} density itself changes discontinuously
near the putative glass transition density for hard sphere systems,
whereas in RFOT theory it is the analogue of the Edwards-Anderson
order parameter Eq.(\ref{eq:order parameter}), which jumps discontinuously
at the ideal glass transition temperature.

The second key idea in the formulation of RFOT is that in general
one expects a very large number of distinct metastable glassy states
\cite{Goldstein_1969} . If the number is large enough this in turn
leads to two distinct transitions. This is indeed what happens in
a many of exactly soluble mean field spin models \cite{Kirkpatrick87PRL,Kirkpatrick87PRB,Kirkpatrick_Wolynes_1987b,Thirumalai88PRB},
in exact high dimensional fluid models \cite{Kurchan_et_al_2012,Kurchan_et_al_2013},
and in mean-field approximations for a variety of liquid state models
\cite{Kirkpatrick_Thirumalai_1989}. In all of these cases the following
scenario is realized. Denote a particular glass state by the label
$\alpha$, with the frozen density in that state given by $n_{\alpha}=n_{l}+\delta n_{\alpha}$
and the free energy in that state equal to $F_{\alpha}$. Below a temperature
denoted by $T_{d}$ ( the $d$ here stands for dynamical, cf. below),
there are an extensive number%
\footnote{An extensive number of states scales like $\exp(c N)$ for a
$N$particle system with $c$ a constant for large $N$.%
} of statistically similar states incongruent states (basically uncorrelated)
that have zero overlap \cite{Huse87JPhysA}: 
\begin{equation}
q_{\alpha\alpha'}=\delta_{\alpha\alpha'}q=\frac{1}{V}\int d\mathbf{x}\delta n_{\alpha}(\mathbf{x})\delta n_{\alpha'}(\mathbf{x})
\end{equation}
 In RFOT it is assumed that 
these features are also realized in realistic structural glass systems with the only caveat that there is no strict dynamical transition at $T_d$. Rather, the transition at $T_d$ is avoided but the dynamics changes around $T_d$ signaling the importance of activated transitions.

Because the states are statistically similar one cannot simply use
an external field to pick out a particular state. The canonical free
energy, $F_{c}$, is given by the partition function via 
\begin{equation}
Z=\exp[-\beta F_{c}]=Tr\exp[-\beta H]=\sum_{\alpha}\exp[-\beta F_{\alpha}].
\end{equation}
 In the glassy context there are two important cases when $F_{c}$
is not the physical free energy. First, if the barrier between the
states is actually infinite then $F_{c}$ is not a physically meaningful
free energy. Second, if the barriers are finite but the experimental
time scale is too short for fluctuations to probe the various states,
then it is also not a physical free energy.

A component averaged free energy can be defined by \cite{Palmer82AdvPhys,Bouchaud_Biroli_2004}
\begin{equation}
\overline{F}=\sum_{\alpha}P_{\alpha}F_{\alpha},
\end{equation}
 with $P_{\alpha}$ the probability to be in the state $\alpha$, 
\begin{equation}
P_{\alpha}=\frac{1}{Z}\exp[-\beta F_{\alpha}].
\end{equation}
 The two free energies, $F_{c}$ and $\overline{F}$, are related
by 
\begin{equation}
F_{c}=\overline{F}+T\sum_{\alpha}P_{\alpha}\ln P_{\alpha}\equiv\overline{F}-TS_{c}.\label{eq:canonical free energy}
\end{equation}
 Here $S_{c}$, is the configurational entropy (sometimes called the complexity or state entropy),
introduced in Section I.B. In general $S_{c}$ is related to the
solution degeneracy and is extensive (and $F_{c}\neq\overline{F}$)
if there are an exponentially large number of states. Note that in
infinite range models with a RFOT and a nonzero $S_{c}$ the physical
free energy is $\overline{F}$ because the $S_{c}$ in Eq. (\ref{eq:canonical free energy})
is an entropy term which is a measure of parts of state space not probed in a finite amount of time. Since a physical entropy
should only be associated with accessible configurations it follows
that $F_{c}$ is not a physically meaningful free energy.

The scenario for the two transitions in the RFOT theory can be described
as follows. For $T>T_{d}$ transport is largely not collective, and
the topology of state space is unremarkable. However, for $T\rightarrow T_{d}$
the dynamics slows down and the system gets stuck in a glassy metastable state. For $T<T_{d}$, there are an extensive
number of statistically similar, incongruent globally glassy metastable
states. If activated transport is neglected these states are infinitely
long lived. The liquid state free energy, $F_{l}$ is lower than the
physical glassy state free energy, $\overline{F}$, but it is equal
to the canonical free energy, $F_{c}$. Because there are so many
glassy states, a liquid with probability one will be stuck in one
of the metastable glassy states for $T<T_{d}$. In the absence of
activated transport it will remain in that state forever. For infinite
range models with an RFOT, exact dynamical calculations shows a continuous
slowing down and freezing as $T\rightarrow T_{d}^{+}$. The same result
is also found for some approximate, mean-field like, calculations
of dynamics in realistic liquid state models. The transition at $T_{d}$
is also closely related to the so-called mode coupling theory of the
glass transition. In realistic systems, activated transport does take
place, and hence on the longest time scales for, $T<T_{d}$, the dynamics
are very sluggish. For this reason $T_{d}$ is called a dynamical
transition: It is a sharp transition only in infinite range models,
but in general it sets a temperature region where the dynamics becomes
glassy like. In addition, for $T<T_{d}$ dynamic heterogeneity (DH)
plays an increasing important role. This too, can be explained as
arising from the multiplicity of states, as we show below.

The driving force for the activated transport in the RFOT scenario
for $T<T_{d}$ is entropic and is given by the state or configurational
entropy. At a lower temperature denoted by $T_{K}$, after the so-called
Kauzmann temperature, the configurational entropy vanishes as does
activated transport. In other words, there is a second transition
at $T_{K}$ which is the ideal or equilibrium glass transition temperature.
For hard sphere systems for which the volume fraction ($\phi$) is
the relevant variable the analogues of the two transitions and emergence
of other phases as $\phi$ is increased is schematically shown in
Fig.(\ref{Francesco1Fig2}).

\subsection{The dynamical transition}

\subsubsection{Theoretical description}

\label{subsec:II.A} The dynamical transition is characterized by
the order parameter, 
\begin{equation}
q(\mathbf{x-y},t)=<\hat{q}(\mathbf{x,y},t)>=<\delta n(\mathbf{x},t)\delta n(\mathbf{y},0)>.
\end{equation}
 Above $T_{d}$ this correlation function for fixed $\mathbf{x-y}$
decays as $t\rightarrow\infty$ but as $T\rightarrow T_{d}^{+}$ its
decay gets slower and slower in a power law fashion. The spatial Fourier
transform of this quantity is the intermediate scattering function,
\begin{equation}
q(k,t)=<F_{\mathbf{k}}(t)>
\end{equation}
 with 
\begin{equation}
F_{\mathbf{k}}(t)=\frac{1}{V}\sum_{i=1}e^{i\mathbf{k\cdot r_{i}(t)}}\sum_{j=1}e^{-i\mathbf{k\cdot r_{j}(0)}}.\label{eq:dynamic order parameter}
\end{equation}
 Here $\mathbf{r}_{i}(t)$ is the position of particle $i$ at time
$t$ and $V$ is the system volume.

We illustrate the dynamical behavior of $q(k,t)$ in Fig. (\ref{KangFig3}a)
as a function of $t$ for a wavenumber equal to the first peak in
the static structure factor, $k_{max}$. The data are taken from \cite{Kang_Kirkpatrick_Thirumalai_2013}
a Brownian dynamics simulation of a binary mixture of highly charged
spherical colloidal particles, a system that becomes a Wigner glass
\cite{Lindsay82JCP} when the fraction of colloidal particles, $\phi$,
increases beyond $\phi_{d}$ the analogue of $T_{d}$.
In the RFOT description of this system, there is a dynamical transition
packing fraction, $\phi_{d}$, as well as an ideal glass transition
value, $\phi_{K}$, with $\phi_{d}<\phi_{g}<\phi_{K}$. Because of
equilibration problems, the simulations are restricted to $\phi$
values just past a packing fraction interpreted as $\phi_{d}$ because
it exhibits all the characteristics of the dynamical transition density
for colloidal particles \cite{Kang_Kirkpatrick_Thirumalai_2013}.
The plots show that in the liquid phase $<F_{\mathbf{k}}(t)>$ decays
for long times, but as the dynamical transition is approached the
system becomes sluggish and $<F_{\mathbf{k}}(t)>$ plateaus for longer
and longer times.

A fundamental quantity of interest is, 
\begin{equation}
q(t)=\frac{1}{V}\sum_{\mathbf{k}}<F_{\mathbf{k}}(t)>
\end{equation}
 At $T_{d}$, $q(t)$ no longer decays, except (in non-mean-field
models) on the longest times scales. Effectively, it becomes the Edwards-Anderson
order parameter for the glass transition: 
\begin{equation}
q_{EA}=\lim_{t\rightarrow\infty}q(t)
\end{equation}
 In RFOT this dynamical order parameter is identical to the equilibrium
$q$ given by Eq.(\ref{eq:order parameter}). In other words, equilibrium-like
theories are in accord with dynamical theories.

Because the OP involves the square of the density fluctuations it
is clear that the associated susceptibility will be something like,
\begin{equation}
\chi_{OP}(\mathbf{x}-\mathbf{y},t)=<\hat{q}(\mathbf{x},\mathbf{y},t)\hat{q}(\mathbf{\mathbf{y}},\mathbf{x},t)>-<\hat{q}(\mathbf{x},\mathbf{y},t)>^{2}
\end{equation}
 Indeed, in \cite{Kirkpatrick_Thirumalai_1988b} the analog of this
quantity was shown to be the relevant susceptibility at a spin glass
transition in the RFOT universality class. Also of interest is the
spatial and time Fourier transform of $\chi_{OP}(\mathbf{x}-\mathbf{y},t)$,
\begin{equation}
\chi_{OP}(k,\omega)=\int dt\int d\mathbf{x}\exp[-i\mathbf{k\cdot(x}-\mathbf{y})+i\omega t)]\chi_{OP}(\mathbf{x}-\mathbf{y},t)
\end{equation}
 A related susceptibility%
\footnote{The homogeneous order parameter susceptibility is given by $\chi_{OP}(0,t)\sim\sum_{\mathbf{k_{1}k_{2}}}[<F_{k_{1}}(t)F_{k_{2}}(t)>-<F_{k_{1}}(t)><F_{k2}(t)>].$
For long times this becomes the wavenumber integral of $\chi_{4|F_{k}(t)}$.%
} that is easily measured in simulations has also been defined, 
\begin{equation}
\chi_{4|F_{k}}(t)=\frac{1}{V}\left[<F_{k}^{2}(t)>-<F_{k}(t)>^{2}\right].\label{eq:chi-4}
\end{equation}
 In Fig.(\ref{KangFig3}b) we show simulation results for this quantity
for the same \cite{Kang_Kirkpatrick_Thirumalai_2013} Wigner glass
as the dynamical transition is approached. In general, one finds that
the location of the maximum and its amplitude grows as the dynamical
transition is approached. This correlation function will be further
discussed in Section III.

The 'static' susceptibility for the glass transition at $T_{d}$ is
\cite{Kirkpatrick_Thirumalai_1988b} 
\begin{equation}
\chi_{OP}(k)=\lim_{\omega\rightarrow0}\chi_{OP}(k,\omega).
\end{equation}
 As $r=T/T_{d}-1\rightarrow0$, the homogeneous susceptibility, $\chi_{OP}=\chi_{OP}(k\rightarrow0)$,
in a mean-field theory, diverges as \cite{Kirkpatrick_Thirumalai_1988b},
\begin{equation}
\chi_{OP}\sim1/\sqrt{r}.
\end{equation}
 At finite and small wavenumber, on the other hand, 
\begin{equation}
\chi_{OP}(k)\sim1/(k^{2}+\xi_{o}^{-2}\sqrt{r}).
\end{equation}
 with $\xi_{o}$ a microscopic (correlation) length. This defines
a divergent length scale as $T\rightarrow T_{d}^{+}$ given by \cite{Kirkpatrick_Wolynes_1987b,Kirkpatrick_Thirumalai_1988b}
, 
\begin{equation}
\xi\sim\xi_{o}/r^{1/4}
\end{equation}
 This is the same divergence found at a mean-field spinoidol point.
It is worth emphasizing again that all of these so-called dynamical
results, also follow from the equilibrium theory of the RFOT. It is important to note that they are mean field results for a phase transition that is avoided in realistic systems. Thus, the predicted exponents are effective exponents.

\subsubsection{Experimental evidence for the dynamical transition}

 Although there is considerable debate about the existence of $T_k \sim T_0$ there is compelling evidence that the very nature of transport in liquids changes at $T_d > T_g$. In an insightful paper, Goldstein argued forty five years ago that the crossover from liquid like dynamics to transport that involves overcoming free energy barriers \cite{Goldstein_1969} occurs at temperatures that far exceed $T_g$. He predicted that barrier crossing events start to become important as soon the relaxation time exceeds 10$^{-9}$s. More recently, by analyzing experimental data of a number of glass forming materials it has been shown that the crossover time is approximately $\tau_c \approx$10$^{-7}$s \cite{Novikov03PRE}. Remarkably, it was noted that $\tau_c$ might be semi-universal. The material-dependent ratio $\frac{T_d}{T_g}$ ranges from 1.1 - 1.7 (see Table 1 in \cite{Novikov03PRE}). As we already pointed out multiple times, $T_d$ roughly corresponds to the temperature predicted by the MCT with the caveat that at $T_d$ the power law singularity describing the dependence of the relaxation time on temperature is avoided in glass forming materials.  These crucial studies demonstrate the relevance of $T_d$ in systems without quenched disorder, which invalidate the strict claims made based on the $p =3$-spin glass model in finite dimensions \cite{Moore02PRL}.

\subsection{Dynamics and random field effects below the dynamical transition}

To a limited extent fluctuations about the RFOT for the SGT have been
considered. From our viewpoint the most important aspect of this work
has been the establishment of the connection between the SGT problem
and the random magnetic field one \cite{Biroli_et_al_2013,Franz_et_al_2012,Franz_et_al_2013}
.

The crucial physical point to see this connection, is to realize that
for $T<T_{d}$ ergodicity is broken on all but the longest time scales.
This immediately implies that there is a difference between averages
over trajectories and averages over initial conditions. In the non-ergodic
phase, if time averages over trajectories are first performed than
the second average over different initial conditions is analogous
to a quenched disorder average. In other words, in the structural glass problem,
in the non-ergodic phase, there is self-induced quenched disorder.
The local overlap function for two different initial conditions or
states $(\alpha,\beta)$ is, 
\begin{equation}
Q_{\alpha\beta}(\mathbf{x})=n_{\alpha}(\mathbf{x})n_{\beta}(\mathbf{x})
\end{equation}
 with $n_{\alpha}(\mathbf{x})$ the density in state $\alpha$. In
the high temperature phase the saddle-point solution is $ $$Q_{\alpha\beta}^{sp}=C$
for all $\alpha\neq\beta$. For $T\leq T_{d}$ there is a simple type
of 'replica' symmetry breaking. The order parameter fluctuation $\phi_{\alpha\beta}=Q_{\alpha\beta}-\mathrm{C}$
satisfies the action \cite{Franz_et_al_2012}, 
\begin{equation}
S=S_{2}+S_{3}+S_{4}\cdots\label{eq: RF STG theory}
\end{equation}
 with $S_{j}$ of $O(\phi^{j}).$ The first few terms are, 
\[
S_{2}=\frac{1}{2}\int d\mathbf{x[\sum_{\alpha\beta}(\nabla\phi_{\alpha\beta})^{2}+\mathit{m_{1}}\sum_{\alpha\beta}\phi_{\alpha\beta}^{2}}
\]
 
\begin{equation}
+m_{2}\sum_{\alpha}(\sum_{\beta}\phi_{\alpha\beta})^{2}+m_{3}(\sum_{\alpha\beta}\phi_{\alpha\beta})^{2}],
\end{equation}
 
\begin{equation}
S_{3}=-\frac{1}{6}\int d\mathbf{x}\left(w_{1}tr\phi^{3}+w_{2}\sum_{\alpha\beta}\phi_{\alpha\beta}^{3}\right),
\end{equation}
 and 
\begin{equation}
S_{4}=\frac{1}{4}\int d\mathbf{x}\left(u_{1}tr\phi^{4}+\cdots\right).
\end{equation}
 The mean-field dynamical transition occurs at $m_{1}=0$. It is easily
seen that the random field aspects of this action is reflected by
the $m_{2}$ and $m_{3}$ terms in $S_{2}$. In particular, to leading
singular order all the Gaussian propagators behave as, 
\begin{equation}
<\phi_{\alpha\beta}(\mathbf{k})\phi_{\gamma\delta}(-\mathbf{k})>=\frac{-4(m_{2}+m_{3})}{(k^{2}+m_{1})^{2}}+\cdots
\end{equation}
 That is, it is proportional to a propagator squared, which is characteristic
of a random field problem \cite{Nattermann_1997,Young_1998} .

The theory given by Eq. (\ref{eq: RF STG theory}) is a random field
problem with cubic terms, which in general reflects the symmetry difference
between fluids and most random magnet problems. Interestingly, it
has been shown that this theory has a critical point when $w_{1}=w_{2}$
where the cubic term vanishes \cite{Franz_et_al_2012}.

The implications of this connection of the SGT problem to the random
field problem is not completely understood. The arguments are perturbative
in nature and seem to neglect the activated transport that takes place
for $T<T_{d}$. Nevertheless, one speculates that some of the features
important in random field magnets such as activated scaling are also
relevant in the SGT problem.

\subsection{Entropy crisis and divergent activated transport near the ideal glass
transition}

\label{subsec:II.C} The RFOT is a new type of discontinuous phase
transition. General arguments and exact calculations in infinite range
models indicate that a divergent coherence or correlation length exist
as an ideal glass transition is approached. Physically, this divergent
length is like a finite size scaling length. Generalizing arguments
of Fisher and Berker \cite{Fisher_Berker_1982} for regular first
order phase transition to a RFOT gives the correlation length that diverges as
\footnote{We use $r=T/T_{d}-1$ as the dimensionless distance from the dynamical
transition and $\epsilon=T/T_{K}-1$ as the dimensionless distance
from the ideal glass transition.}
 $\xi\sim 1/\epsilon^{\nu}$, with $\epsilon=T/T_{K}-1$, and correlation length exponent
\cite{Kirkpatrick_Thirumalai_Wolynes_1989}, 
\begin{equation}
\nu=\frac{2}{d}.
\end{equation}
 This result is expected \cite{Kirkpatrick_Thirumalai_2014} to be exact for all dimensions where a transition takes place. An exact finite size scaling calculation \cite{Kirkpatrick_Wolynes_1987b}
for an infinite range model undergoing an RFOT also gives this value.
In this case
 as $\epsilon\rightarrow0$ the configurational entropy
per site vanishes as, 
\begin{equation}
S_{c}/N\sim\epsilon
\end{equation}
 and the finite size correlation length diverges as, 
\begin{equation}
\xi\sim1/|\epsilon|^{2/d}
\end{equation}

An important characteristic of a glass transition is the occurrence
of extremely long relaxation time scales. While critical slowing down
at an ordinary transition means that the critical time scale grows
as a power of the correlation length, $\tau\sim\xi^{z}$ with $z$
the dynamical scaling exponent, at a glass transition the critical
time scale grows exponentially with $\xi$, 
\begin{equation}
\ln(\tau/\tau_{o})\sim\xi^{\psi}\label{eq: activated scaling}
\end{equation}
 with $\tau_{o}$ a microscopic time scale, and $\psi$ a generalized
dynamical scaling exponent. Effectively, Eq.(\ref{eq: activated scaling})
implies $z=\infty$. As a result of such extreme slowing down, the
system's equilibrium behavior near the transition becomes inaccessible
for all practical purposes. Thus, realizable experimental time scales
are not sufficient to reach equilibrium, and one says the system falls
out of equilibrium.

Activated scaling, as described by Eq. (\ref{eq: activated scaling}),
follows from a barrier picture of the system's free energy landscape.
In the context of the structural glass transition it is called the
so-called mosaic picture \cite{Kirkpatrick_Thirumalai_Wolynes_1989,Bouchaud_Biroli_2004}.
The basic idea is that for $T_{d}>T>T_{K}$ there is an entropic driving
force that causes a compact, glassy state of size $\xi^{d}$ to make
a transition to a different glassy state, with the same approximate
free energy, also of size $\xi^{d}$. The physical picture that results
is a system that looks like a mosaic, or patchwork, of different glassy
regions separated by diffuse or fuzzy interfaces slowly making a transition
to yet other glassy states. For the uncorrelated states that exist
above $T_{K}$ the law of large numbers \cite{Thirumalai_Mountain_Kirkpatrick_1989}
is consistent with a barrier that scales like $\sim\xi^{d/2}$. This
is also consistent with scaling and an entropic driving force $\epsilon\xi^{d}\sim\epsilon^{-1}$,
if $\nu=2/d$. In the original RFOT paper \cite{Kirkpatrick_Thirumalai_Wolynes_1989},
a wetting argument, along the lines proposed for RFIM \cite{Villain_1985},
was given that also led to barriers scaling like $\xi^{d/2}$. All
of this in turn implies a Vogel-Fulcher law for the temperature dependence
of the relaxation time should hold as the glass transition is approached:
\begin{equation}
\tau\sim\tau_{o}\exp[\frac{D}{T/T_{K}-1}]
\end{equation}
 with $D$ a positive constant. Within RFOT and related theories $T_{K}$ is identified with $T_{0}$ in the VTF equation, Eq.(\ref{VFT}).

As noted in Section I.D that the barriers scaling like $\xi^{d/2}$ implies there is
really no interface between two statistical similar glassy states.

\section{Dynamic heterogeneity, law of large numbers, rare regions, and activated
scaling in glassy systems}

\subsection{Dynamic heterogeneity}

Experimentally, various spectroscopic techniques have revealed $\mathit{hetrogeneous}$
relaxation in glassy systems \cite{Bouchaud_Biroli_2005,Berthier_2011,Ediger_2000}
. In such systems, there is non-exponential decay of correlations
that can be explained as arising from the superposition of different
regions decaying with different relaxation rates.

A large number of molecular simulations have provided visualization
of the microscopic details of the dynamical heterogenities in glass
forming systems \cite{Donati_et_al_2002,Ediger_Harrowell_2012}. These
simulations have provided direct evidence of dynamic heterogeneities,
i.e., the existence of finite time correlated domains with a length
scale that can exceed the molecular scale. An illustrative simulation
result is shown in Fig. (\ref{DHFig4}). Experiments to directly visualize
these dynamic heterogeneities have also been performed in colloidal
glasses.

The chief theoretical construct used to understand dynamic heterogeneity
near the dynamical transition is the order parameter susceptibility
and the related function $\chi_{4|F_{k}}(t)$, both of which are defined
in Section II.B. In Fig.(\ref{KangFig3}b), we show $\chi_{4|F_{k}}(t)$
as a function of $t$ for a wavenumber equal to that of the first
peak of the static structure factor, $k_{max}$ for a system forming
a Wigner glass. In general, one finds that there is a peak that becomes
larger and moves to longer times as the glass transition is approached.

Ordinary scaling ideas can be used to partially explain and interpret
these results. The simulations show that the location in the peak
of $\chi_{4|F_{k}}(t)$ increases as a power law as $\phi_{d}$ is
approached from below according to \cite{Kang_Kirkpatrick_Thirumalai_2013},
\begin{equation}
t^{*}\sim(\phi^{-1}-\phi_{d}^{-1})^{-\gamma_{\chi}}\label{eq:peak location chi-4}
\end{equation}
 
\begin{equation}
\gamma_{\chi}\simeq1.05\label{eq:peak location exponent}
\end{equation}
 General scaling ideas, on the other hand, give that $t^{*}\sim\xi^{z}\sim1/r^{\nu z}$
with $z$ the dynamical scaling exponent, $r=(\phi^{-1}-\phi_{d}^{-1})$
is the distance from the dynamical transition, and $\nu$ is the correlation
length exponent for the dynamical transition. According to the mean-field
description in Section II.B, $\nu=1/4$. This along with Eqs. (\ref{eq:peak location chi-4})
and (\ref{eq:peak location exponent}) gives $z\simeq4.2.$This result
is consistent with the results of Kim and Saito \cite{Kim_Saito_2013}
.

In non-equilibrium aging simulations and experiments there is a growing,
time dependent, correlation length that has been measured. If we assume
ordinary scaling (also see, Section III.D below) to describe the $r$
and $t$ dependence of the correlation length then the natural assumption
is, 
\begin{equation}
\xi(r,1/t)=bf_{\xi}(b^{1/\nu}r,b^{z}/t)
\end{equation}
 with $b$ an arbitrary length rescaling factor and $f_{\xi}$ a scaling
function. Choosing $b=t^{1/z}$ gives, 
\begin{equation}
\xi(r,1/t)=t^{1/z}f_{\xi}(rt^{1/\nu z},1)
\end{equation}
 This implies that for $rt^{1/\nu z}<1$ there is a correlation length
that grows in time as $\sim t^{1/z}\sim t^{.24}$.

\subsection{Dynamic heterogeneity and violation of law of large numbers}

Ingenious four-dimensional NMR experiments \cite{Sillescu_1999,Sillescu_et_al_2002}
and dielectric relaxation measurements have provided the needed evidence
for heterogeneous dynamics in glass forming materials. However, much
of our understanding of the notion of DH comes from computer simulations,
most of which have been quantified using the four-point dynamic susceptibility
function. The lack of symmetry breaking as the SGT occurs forces us
to use higher order correlation functions to distinguish between the
liquid and the glassy phase. Within the RFOT (and MCT) formalism the
natural dynamic order parameter is the two-point intermediate scattering
function given by Eq. (\ref{eq:dynamic order parameter}). It decays
to zero in the liquid phase, and acquires a plateau whose duration
grows as the extent of supercooling increases Fig.(\ref{KangFig3}a).
Thus, it is necessary to use the fluctuations in $F_{k}(t)$, which
plays the role of generalized susceptibility, $\chi_{4|F_{k}(t)}$,
Eq.(\ref{eq:chi-4}), to distinguish between the states accessed above
and below $T_{d}$ \cite{Kirkpatrick_Thirumalai_1988b} . Although
it is physically most meaningful to use fluctuations in $F_{k}(t)$
a number of studies have used $\chi_{4|S}$ ($S$ is some observable)
to infer the nature of dynamical heterogeneity in several model systems
\cite{Toninelli_et_al_2005,Dasgupta_et_al_1991,Donati_et_al_2002,Bouchaud_Biroli_2005}
. The four point correlation function $\chi_{4|F_{k}(t)}$ ($S=F_{k}$)
is the variance in $F_{k}\left(t\right)$. For a large number of systems
it is found that $\chi_{4|F_{k}}(t)$, at a specified $k$, has a
peak in the time-domain with the amplitude that grows with increased
supercooling. In Fig.(\ref{KangFig3}b), we show a typical dependence
of $\chi_{4|F_{k}}(t)$ at $k_{max}$ for a Wigner glass for which
increasing volume fraction of the colloidal particles is roughly analogous
to decrease in temperature. By computing the $k$-dependence of $\chi_{4|F_{k}}(t)$
an estimate for the length scale, $\xi_{DH}$, associated with dynamic
heterogeneity (DH length) can be made with the assumption that the
maximum amplitude $\chi_{4|F_{k}}(t)$ follows the Ornstein-Zernicke
form, $\chi_{4|k}(t)\approx\frac{\xi_{DH}^{2}}{(1+(\xi_{DH}k)^{2})}$.

A physical consequence of the length scale associated with dynamic
heterogeneity is that the usual law of large numbers, which is obeyed
in liquids, is violated in the glassy phase \cite{Thirumalai_Mountain_Kirkpatrick_1989}.
The plausible emergence of a natural length scale within which the
particles are highly correlated allows us to imagine that the below
$T<T_{d}$ the entire sample can be partitioned into subsamples whose
size can be associated with the DH length. As the temperature decreases
we expect this length to be large enough that meaningful averages
over the number of particle within DH length can be performed. In
the liquid phase ($T<T_{d}$) the statistical properties of the liquid
(for example the average energy of particles of a given type) would
be independent of the subsample size, and should will coincide with
that of the entire sample (within the usual fluctuation effects) provided
the DH length is large. This is the usual statement that the law of
large numbers is expected to hold in the ergodic liquid phase. On
the other hand, in the glassy phase each subsample is likely to be
distinct, and consequently there ought to be variations between one
subsample to another. Because the time for rearrangement of one subsample
to another gets slower and slower as the degree of supercooling increases
the in-equivalence between particles of a given type between two samples
would persist even on the observation time, $\tau_{obs}$. Thus, no
single sub sample can statistically characterize characterize the
equilibrium properties of the entire sample, even after suitable time
average . In other words, in the glassy phase the law of large numbers
is violated, and there are ought to be subsample to subsample fluctuations.
Only by examining the entire sample on timescales that far exceed
the observation times can these intrinsic heterogeneities between
subsamples become irrelevant. This physical picture suggests that
DH dynamical heterogeneity a consequence of the emergence of glassy
clusters, which are essentially frozen with relaxation time that far
exceeds $\tau_{obs}$. Because of the variations in both equilibrium
and relaxation properties from subsample to subsample a glassy phase
is inherently heterogeneous, as noted in several studies.

These concepts were illustrated using computer simulations of binary
soft sphere mixtures \cite{Thirumalai_Mountain_Kirkpatrick_1989},
and more recently mixtures of charges colloidal particles, which form
Wigner glasses at high densities or volume fractions \cite{Kang_Kirkpatrick_Thirumalai_2013}.
This experimentally characterized system is liquid-like at volume
fractions $\phi$ below $\phi\approx\phi_{d}\approx$ 0.1, and turns
into a Wigner glass above $\phi>.1$. We divided the simulation sample
into subsamples with appropriate size determined by an approximate
measure of structural entropy. It order to establish the violation
of law of large numbers we showed in Fig.(7) of \cite{Kang_Kirkpatrick_Thirumalai_2013}
the time evolutions of distribution of the structural entropy, $s_{3}$
for a large subsample and the whole sample for $\phi=0.02$ and $\phi=0.2$.
As expected based on law of large numbers, we found that in the liquid
phase ($\phi=0.02$) the distributions $P(\bar{s_{3}}|t)$ are almost
the same for all $t$ values that exceed the typical relaxation time.
In contrast, at higher volume fractions ($>\phi_{d}$) where ergodicity
is effectively broken, the $P(\bar{s_{3}}|t)$ for the subsample are
substantially different from that of the entire sample, thus violating
the law of large numbers . Because different subsamples behave in
a distinct manner and do not become equivalent, we surmise that dynamical
heterogeneity is a consequence of violation of law of large numbers.
It should be noted that only by examining the time evolution of the
subsamples in the liquid and the glassy phase can this link be established.
The intuitive arguments given here are made more precise in the following
section.

\subsection{Rare region dynamics near the glass transitions}

The existence of DH suggests that in a very viscous liquid the longest
time decay of any time correlation function will be determined by
the large rare region or anomalous clusters of particles of some linear
dimension $L$ \cite{Berthier_2011} . These large clusters are fluidized
and can relax to a more typical configuration of particles in some
characteristic time $\tau(L)$. For this argument to be sensible $L$
must be larger than a molecular scale. To estimate the effect of these
large rare regions on a typical time correlation function an average
over $L$ must be performed.

Since the large clusters are rare, we assume that their probability
distribution is controlled by Poisson statistics so that the tail
probability of an unusual cluster of size $L$ is%
\footnote{Although related, this assumption is physically distinct from what
is used in quenched disordered systems \cite{Vojta_2006}. Here we
simply postulate that since the events are rare, they are controlled
by a Poisson distribution.%
} , 
\begin{equation}
P(L)\sim\exp(-cL^{d})
\end{equation}
 with $c$ a positive constant. We also assume that a typical correlation
associated with the rare region decays exponentially as, 
\begin{equation}
C(L,t)\sim\exp[-t/\tau(L)].
\end{equation}
 It is also reasonable to assume that the long time dynamics of these
fluidized regions is diffusive so that $1/\tau(L)\rightarrow Dk^{2}\rightarrow D(L)/L^{2}$.
We consider two temperature regions. The first is appropriate for
temperatures near $T_{d}$ and the second for temperatures close to
$T_{K}$ or the laboratory glass transition temperature, $T_{g}$.

In the first region the scale dependence of $D$ is ignored so that
the average correlation function decays as, 
\[
C(t\rightarrow\infty)\sim\int dL\exp[-cL^{d}-Dt/L^{2}]
\]
 
\begin{equation}
\sim\exp[-A(Dt)^{d/(d+2)}]\label{eq:stretched exponential}
\end{equation}
 with $A$ a positive constant. The characteristic length scale $L_{1}^{*}(t)\sim[Dt]^{1/(d+2)}$.
Equation (\ref{eq:stretched exponential}) is the stretched exponential
behavior typically observed in correlation functions in simulations
near $T_{d}$ with a large time, $\tau$, scale given by $\tau\sim1/D$.
For example, the solid lines in Fig.(\ref{KangFig3}a) are a fit to
a stretched exponential $\sim\exp[-(t/\tau)^{\beta}]$ with $\beta=0.45$.
In general a distribution of relaxation times, $P(\tau)$, can be
defined by writing 
\begin{equation}
C(t)\sim\int d\tau P(\tau)\exp[-t/\tau].\label{Stretch2}
\end{equation}
 By comparing Eq. (\ref{eq:stretched exponential}) and Eq. (\ref{Stretch2})
it follows that, 
\begin{equation}
P(\tau\rightarrow\infty)\sim\exp[-(D\tau)^{d/2}],
\end{equation}
 with a tail that decays faster than exponential.

In the second, lower temperature, region, the scale dependence of
the diffusion constant becomes most important. If we assume that $D(L)$is
inversely proportional to the RFOT relaxation time, 
\begin{equation}
\tau(L)=\tau_{m}\exp[aL^{d/2}]
\end{equation}
 with $\tau_{m}$ a microscopic time and $a$ a positive constant.
Using all of this an average correlation function then decays for
long times as, 
\[
C(t)\sim\int dL\exp[-cL^{d}-(t/\tau_{m})(e^{-aL^{d/2}}/L^{2})]
\]
 
\begin{equation}
\sim\exp\{-A[\ln(t/\tau_{m})]^{2}\}
\end{equation}
 with $A$ a positive constant. The conclusion is that for long times
$C(t)$ decays faster than any power law, but slower than any stretched
exponential. The characteristic length scale for this case is $L_{2}^{*}(t)\sim(\ln t)^{2/d}$.

In this case, the distribution of relaxation times, $P(\tau)$, is
given by, 
\begin{equation}
P(\tau\rightarrow\infty)\sim\exp\{-c[\ln(\tau/\tau_{m})]^{2}\}
\end{equation}
 with a characteristic tail that decays slower than any exponential.

Finally we note that in a given system well below $T_{d}$ there will
be an intermediate time region where the scale dependence of $D(L)$
is not important and a stretched time behavior will be observed, before
crossing over to the exponential of $[\ln t]^{2}$at the longest times.
The crossover time will be roughly given by the equation $L_{1}^{*}(t)\sim L_{2}^{*}(t)$.

\subsection{Activated scaling near the glass transition}

Activated scaling was developed to understand finite dimensional (three
dimensions) spin glasses and random field magnets where the dynamics
is controlled by large, possibly divergent, free energy barriers \cite{Fisher_Huse_1988}.
Similar ideas can be applied to the structural glass problem, also
in three-dimensions.

Here, we examine the behavior of the glass transition susceptibility,
introduced in Sec II.B using activated scaling ideas \cite{Fisher_Huse_1988}
as the ideal glass transition is approached. We start with the observation
that the first order nature of the ideal glass transition implies
that the scale dimension of $q(\mathbf{x},t)$ is zero. This and the
activated scaling ansatz gives that the wavenumber and time dependent
glass transition susceptibility will satisfy the scaling law, 
\begin{equation}
\chi_{OP}(\epsilon,k,t)=b^{d}F_{\chi}[\epsilon b^{1/\nu},bk,\frac{b^{d/2}}{\ln(t/t_{o})}]\label{suseq}
\end{equation}
 where $\epsilon=T/T_{K}-1$ is the dimensionless distance from the
ideal glass transition, $t_{o}$ is some microscopic time scale, and
$F_{\chi}$ is a scaling function. Note that we have used here that
the barrier height scales as $b^{d/2}\sim\xi^{d/2}$. This equation
implies a number of non-trivial results. For example, at zero wavenumber,
and at the ideal glass transition temperature we can choose $b=[\ln(t/t_{o})]^{2/d}$
to obtain, 
\begin{equation}
\chi_{OP}(0,0,t\rightarrow\infty)\sim[\ln(t/t_{o})]^{2}.\label{susNL}
\end{equation}
 This dynamic scaling result is valid as long as $\epsilon\ln(t/t_{o})<1$.
It also defines a dynamic crossover $\epsilon$ being given by, 
\begin{equation}
\epsilon_{x}\sim1/[\ln(t/t_{o})]\label{crossover}
\end{equation}
 Physically this means that the large correlations that exist at $T_{K}$
can be measured by examining the slow growth in time of the glass
transition susceptibility around $k=0$. This should be experimentally
relevant. If the exponent of $2$ in Eq.(\ref{susNL}) can be experimentally
demonstrated then it would be very strong evidence for the validity
of the RFOT theory of the SGT.

The frequency dependent glass transition susceptibility defined by
Eq.(\ref{susNL}) can similarly be expressed as a scaling function.
In general the $\epsilon_{x}$ given by Eq.(\ref{crossover}) will
give the scale distinguishing static critical behavior from dynamical
critical behavior for all quantities as $T\rightarrow T_{K}$.

Although not as rigorously founded as the scaling law for $\chi_{NL}$,
we can also give a scaling law for the frequency dependent shear viscosity,
$\eta(\epsilon,\omega)$. Because $\eta(\epsilon,\omega)$ is related
to a time integral of a time correlation function its static value
is proportional to $\tau$ given by Eq. (\ref{eq: activated scaling}).
We then obtain, 
\begin{equation}
\eta(\epsilon,\omega)=\exp(b^{d/2})F_{\eta}[\epsilon b^{1/\nu},\frac{b^{d/2}}{\ln(1/t_{o}\omega)}]\label{eta}
\end{equation}
 with $F_{\eta}$ a scaling function. The static or zero frequency
shear viscosity then behaves as $\tau$ but for $\epsilon<1/\ln(1/t_{o}\omega)$
it behaves as 
\begin{equation}
\eta[\epsilon\ln(1/t_{o}\omega)<1]\sim\frac{1}{t_{o}\omega}.\label{etaglassy}
\end{equation}
 Again, the important physical and experimental point is that $\epsilon_{x}$,
given by Eq.(\ref{crossover}), sets the crossover scale in either
time or frequency ($t\rightarrow1/\omega$) space. Note that $\eta$
being simply proportional to $\tau$ in Eq.(\ref{eta}) is needed
to obtain Eq.(\ref{etaglassy}), which in turn is required for the
proper stress/strain relation in the glassy phase.

Following Section III.A we next use activated scaling ideas to describe
the time-dependent aging correlation length. In this case the natural
assumption is, 
\begin{equation}
\xi(\epsilon,1/t)=bF_{\xi}(b^{1/\nu}\epsilon,e^{b^{d/2}}/t)
\end{equation}
 with $F_{\xi}$ a scaling function. Choosing $b^{d/2}=\ln t$ and
using $\nu=2/d$ gives, 
\begin{equation}
\xi(\epsilon,1/t)=(\ln t)^{2/d}F_{\xi}(\epsilon\ln t,1)
\end{equation}
 Thus, we expect that close to the ideal glass transition and for
$\epsilon\ln t<1,$ there ought to be a correlation length in aging
experiments that grows as $\sim(\ln t)^{2/3}$ in ($d=3$).

\section{Understanding biological problems from the perspective of glass physics}

There are several ways in which concepts in glass physics can be used
to understand many aspects of biological systems. At the cellular
level, on length scale on the order of $\mu$m, functions are carried
out often by several interacting biological molecules. Transport in
eukaryotes, supporting cytoskeletal structures, is powered by ATP-driven
motors. However, in \textit{E. Coli.} all dynamical processes occur
by diffusion. Moreover, the dynamics has to occur in a heterogeneous
crowded environment within a restricted time interval with the upper
bound being the cell doubling time. Therefore, it is likely that the
biological molecules only sample a restricted part of the access conformational
space, which implies that ergodicity could well be broken as in a
liquid undergoing glass transition. On longer length scales, involving
communication between cells, needed in diverse phenomena such as development
and wound healing, there are manifestation of glass like behavior
or at least evidence of highly heterogeneous behavior \cite{Altschuler10Cell,Pelkmans12Science,Gross13CurrBiol}.
This is not entirely surprising because these processes involve collective
movements, which can be sluggish. In particular, in tissues without
gaps between cells there is evidence that the collective dynamics
\cite{Angelini11PNAS}, much like correlated movements of particles
in the glassy state, have many of the hall marks of the SGT \cite{Garrahan11PNAS}.
Here, we use a few examples to illustrate that concepts in glass physics,
which at first glance may seem unrelated to biology, are useful in
providing insights into dynamics in biological systems from nm to
$\mu$m, and beyond.

\label{sec:III}

\subsection{Countable number of structural states in the sequence space of proteins}

An astounding aspect of proteins and RNA is that natural foldable
sequences, whose number is much smaller than all possible sequences,
self-organize themselves spontaneously often without the help of molecular
chaperones \cite{OnuchicARPC97,Schuler08COSB,Thirumalai10ARB,Dill08ARB,Shakhnovich06ChemRev}.
Why are the number of structure forming sequences so small? Answering
this question quantitatively forces us to think in terms of partitioning
of the vast sequence space in terms of disjoint states (just as described
in Section (IIA)). By envisioning the partitioning of both the sequence
and conformational space of proteins and RNA in terms of the associated
landscapes, we can begin to appreciate the emergence of structures
as well as the characteristics of sequences that make them biologically
viable, implying that they fold relatively rapidly.

Here, we only consider proteins. The primary building blocks of proteins
are $\alpha$-helices (one dimensional ordered structures), $\beta$-sheets
(contain two dimensional order), and loops of varying bending rigidity
\cite{Stryer1988}. From these seemingly simple building blocks (referred
to as secondary structural elements) a large number of three dimensional
structures can be constructed. The number of distinct topological
folds is suspected to be only on the order of a few (at best) thousand,-
a relatively small number \cite{Chothia92Nature}. How do these preferred
folds, which should also be kinetically accessible on biologically
relevant time scale, emerge from the dense sequence space? The number
of sequences of a polypeptide chain with $N$ amino acids is $20^{N}$,
which is astronomically large even when $N$ takes on a modest value.
It is likely that only an extremely small fraction of the sequences
encodes for the currently known protein structures. A quantitative
mapping between sequence space and structures, obtained using lattice
models \cite{LiScience96}, shed light on the structure of the sequence
space landscape. In order to appreciate the partitioning of the sequence
space it is worth recalling that natural proteins in their native
states are (i) compact, and (ii) dense interior is made up of predominantly
hydrophobic residues. With these two restrictions on the native structures,
it has been shown that even though the number of sequences is astronomically
large, the number of compact low energy structures (protein-like)
is considerably smaller both in two and three dimensions \cite{Camacho93PRL,Thirumalai99BookChap,Lin12PNAS}.
This would imply that for many sequences the low energy compact structures
could be nearly the same, as was beautifully illustrated by exploring
sequences in a three dimensional lattice model \cite{LiScience96}.
In other words, the basins of attraction in the structure space are
rare enough so that a large number of sequences map on to precisely
one basin, thus explaining the emergence of greatly limited number
of structures from the sea of sequence space \cite{Chothia92Nature}.

Similar considerations hold for RNA with the crucial difference that
RNA structures are lot more degenerate compared to proteins \cite{ThirumARPC01}.
An identical RNA sequence can fold into two distinct structures performing
entirely different functions \cite{BartelSCI00}. This implies that
the sequence space landscape could be multiply connected with larger
number of structurally degenerate states compared to proteins. From
the perspective of navigating the sequence space landscape, which
presumably occurs on the evolutionary time scale, the dynamics is
predicted to be slower than evolvability of protein sequences.

\label{subsec:III.A}

\subsection{Kinetic accessibility and folding rate dependence of proteins and
RNA on $N$}

A corollary of the finding that natural sequences fold into minimum
energy compact strctures quickly is that random sequences cannot exhibit
protein-like behavior both on account of stability and perhaps more
importantly kinetic accessibility of the folded states \cite{Bryngelson89JPC}.
Even if random heteropolymers formed by covalently linking various
amino acids have unique ground states the folding dynamics would be
highly sluggish \cite{Thirumalai96PRL,Takada97PNAS} such that deleterious
aggregation could intervene before folding. A solution to this conundrum
is that the folding transition temperature, $T_{F}$, of sequences
that lead to functional proteins should exceed the equilibrium glass
transition temperature, a suggestion that was based on the extension
of the Random Energy Model (REM) to protein folding with the native
state playing a special role \cite{Bryngelson89JPC}. In the REM,
equivalent to $p$-spin model with $p\rightarrow\infty$, there is
an entropy crisis at a finite temperature. Because of finite size
of proteins there is no strict entropy crisis, and hence it was realized
that $T_{F}$ has to exceed a dynamic glass transition temperature,
$T_{g}$, for folding to the native state to occur in biologically
meaningful time \cite{Socci95JCP}. Ideas based in polymer physics
further showed that the interplay of $T_{F}$, and the equilibrium
collapse temperature ($T_{\theta}$) \cite{Camacho93PNAS} could be
used to not only fully characterize the phase diagram of generic protein
sequences but also determine their foldability, a prediction that
has been experimentally validated only very recently \cite{Hofmann12PNAS}.
Based on the study of dynamics of random copolymer models it was proposed
that the upper bound on $\frac{T_{F}}{T_{gd}}$ is $\frac{T_{F}}{T_{\theta}}$
\cite{Thirumalai96PRL}. Thus, by studying disordered systems exhibiting
glassy behavior insights into foldable sequences were obtained.

The description of activated dynamics using RFOT was also adopted
to obtain an estimate of the dependence of the folding rates of globular
proteins on $N$. The folding reaction typically involves crossing
a free energy barrier, and hence the folding time is given by $\tau_{F}=\tau_{0}e^{\Delta F^{\ddagger}/k_{B}T}$
where $\Delta F^{\ddagger}$ is the average free energy separating
the native state from an ensemble of partially structured and compact
states. The scaling of $\Delta F^{\ddagger}$ with $N$ parallels
the arguments developed in the context of activated dynamics using
RFOT concepts \cite{Thirum95JPI}. We assume that the free energy distribution
of the low energy structures is given by a Gaussian distribution,
which is also consistent with computations on model glass forming
systems. Since there is an ensemble of independent transition states
connecting the conformations of compact but non-native states and
the native state it is natural to assume that the barrier height distribution
is also roughly Gaussian with a dispersion $\langle\Delta F^{2}\rangle$
that scales as $N$. Since the barrier height distribution is essentially
a Gaussian it follows that $\Delta F^{\ddagger}\approx\langle\Delta F^{2}\rangle^{1/2}\approx\sqrt{N}$.
This physically motivated argument is also consistent with the tenets
of RFOT. For proteins the appropriate length scale is essentially the whole protein molecule, and hence 
\begin{equation}
\Delta F^{\ddagger}\approx N^{\frac{1}{\nu d}},\label{barrier}
\end{equation}
 and with $\nu=2/d$ we obtain the important result that the barrier
to folding scales sub linearly with $N$. This scaling type relation
has been successfully applied to rationalize the folding rates of
a large number of proteins whose folding rates cover seven orders of magnitude
(Fig. \ref{ScalingFig5}a).

Although Eq.(\ref{barrier}) explains the data for proteins, we expect
that the theory should account for the folding rate changes with the
number of nucleotides ($N$) even better because RNA has multiple
folded metastable states \cite{Thirum05Biochem}, which could be thought
of as free energy excitations around the native state. It is also
likely that even the functional state for RNA may not be unique \cite{Solomatin10Nature},
thus %
\footnote{There is evidence that in some cases it is likely that even in proteins
the folded state may be metastable, especially in the case of mammalian
prions \cite{Thirumalai03COSB,Baskakov01JBC}%
} making the energy landscape very much glass like. In accord with
this expectation, it has been found that the folding dynamics is sluggish
with trapping in metastable states occurring with high probability
\cite{Pan97JMB}. As a consequence of the highly rough free energy
surface the folding rates can be predicted using Eq. \ref{barrier}.
Remarkably, the folding rates of RNA also obeys $\tau_{F}=\tau_{0}e^{\sqrt{N}}$
with high accuracy with $\tau_{0}\approx1\mu$s (Fig. \ref{ScalingFig5}b).

\label{subsec:III.B}

\subsection{Persistent heterogeneity}

The underlying energy landscapes of biological molecules, especially
large RNA, are rugged consisting of multiple states that are separated
by large barriers. As a consequence, it is most likely the case that
they should exhibit glass-like behavior, which has only been recently revealed
most clearly using single molecule experiments although pioneering
experiments by Frauenfelder \cite{Frauenfelder1988ARBBC} had already
anticipated these possibilities. An important consequence of several
studies is that the functionally competent states of RNA and possibly
proteins may not be unique, as is generally assumed. In terms of the
RFOT description of glasses it implies that there are
many components or states in the folding landscape and just as in
glasses the canonical free energy is not relevant as would be the
case if the folded state did always correspond to the global free
energy minimum \cite{Anfinsen75APC}. The widely accepted notion that
the native state of proteins and RNA are unique was inferred using
bulk ensemble experiments tacitly assuming that ergodicity is established
on $\tau_{obs}$. In a rugged landscape, a specific molecule with
an initial conformation distinct from others only samples limited
conformational space corresponding to a single state. Ergodically
sampling all states would only be possible on time scales longer than
biologically relevant times. This scenario results in heterogeneous
dynamics as in glasses, and ensemble average would obscure the complexity
of the structural features of the underlying landscape. Indeed, recent
findings from single molecule experiments on several biomolecular
systems explicitly showed persistent heterogeneities in time traces
(or molecule-to-molecule variations) generated under identical folding
conditions \cite{ZhuangSCI02,Ditzler08NAR,Okumus04BJ,Solomatin10Nature,Borman10CEN}.
Unlike phenotypic cell-to-cell variability among genetically identical
cells, which can be visualized using microscope \cite{Pelkmans12Science},
the observation of heterogeneity among individual biomolecules on
much smaller length scales is tantalizing because it would make it
difficult to reconcile this concept with the conventional notion that
functional states of proteins and RNAs are unique or that various
native basins of attraction easily interconvert on the time scale
of observation. For example, in docking-undocking transitions of surface
immobilized hairpin ribozyme \cite{ZhuangSCI02} and \textit{Tetrahymena}
group I intron ribozyme \cite{Solomatin10Nature}, time traces for
individual molecules display very different dynamic pattern with long
memory without apparent compromise in catalytic efficiency. Based
on these observations it was suggested that these ribozymes have multiple
native states \cite{Solomatin10Nature}. If this were the case then
it follows from the analogies to glasses that (1) the underlying folding
landscape must contain multiple discernible states with little possibility
of interconversion among them on $\tau_{obs}$ implying that ergodicity
is effectively broken; (2) the dynamics within each state or basin
of attraction ought to be different, which would be a manifestation
of dynamic heterogeneity. Demonstrating these important aspects of
molecule-to-molecule variations resulting in persistent DH using ensemble
experiments is difficult. However, single molecule experiments analyzed
using glass physics concepts have recently have shown that these conclusions
are indeed valid. We use two completely unrelated examples to illustrate
the concept persistent heterogeneity in biological systems at a molecular
scale.

\textit{Holliday junctions:} Holliday Junctions (HJs) are essential
intermediates for strand exchange (Fig. \ref{HJFig6}a) \cite{Lushnikov03JBC}
in DNA recombination. HJs exist in two distinct isoforms (\emph{isoI}
and \emph{isoII}) both of which have the characteristic X-shaped architectures
at high Mg$^{2+}$ concentrations ($\sim$ 50 $\mu$M). Using smFRET
experiments \cite{Schuler08COSB}, and concepts from glass physics
\cite{ThirumalaiPRA89,Kirk89JPhysA} and complementary clustering
algorithms \cite{Tamayo99PNAS,Sturn02Bioinformatics} the state space
structure and associated dynamics were quantitatively analyzed \cite{Hyeon12NatChem}.
Although the HJ dynamics at the ensemble level shows fluctuations
between only two-states, trajectories from smFRET reveal a much richer
structure with the associated dynamics exhibiting some of the hallmarks
associated with glasses. In smFRET experiments efficiency of energy
transfer as a function of $t$ is calculated from the measured donor
(D) ($I_{D,i}(t)$) and acceptor ($I_{A,i}(t)$) emission intensities
as, $E_{i}(t)=I_{A,i}(t)/(I_{A,i}(t)+I_{D,i}(t))$. Thus, smFRET experiments
provided time-dependent ``trajectories\textquotedbl{} in terms of
the collective variable, $E_{i}(t)$, for the $i^{th}$ molecule (Fig.
\ref{HJFig6}b). When an ensemble average over a sufficiently large
number of molecules and time $\tau_{obs}\approx40$ sec is performed
one observes a simple two-state behavior (right side of Fig. \ref{HJFig6}b).

However, detailed analysis of the FRET trajectories revealed surprising
evidence of DH. For a given time trace corresponding to a specific
molecule $\alpha$, $\mathcal{T}_{obs}\approx40$ sec is long enough
to observe multiple isomerization events between \emph{isoI} and \emph{isoII}
the conformations \cite{Hyeon12NatChem}. The time scale for single
isomerization between \emph{isoI} and \emph{isoII} ($\tau_{\mathrm{I\leftrightarrow II}}^{\alpha}$)
is much smaller than $\mathcal{T}_{obs}$ ($\tau_{\mathrm{I\leftrightarrow II}}^{\alpha}\ll\mathcal{T}_{obs}$)
(Fig. \ref{HJFig6}a and Fig. \ref{HJFig7}a). Thus, HJ explores the
conformations in only the $\alpha$ state exhaustively as shown by
the ergodic measure in the upper part of Fig. \ref{HJFig7}b); however
it is not long enough for interconversion to take place between molecules
$\alpha$ and $\beta$, i.e., $\mathcal{T}_{obs}\ll\tau_{conv}^{\alpha\leftrightarrow\beta}$
where $\tau_{conv}^{\alpha\leftrightarrow\beta}$ is the interconversion
time between $\alpha$ and $\beta$ states, implying that a substantially
high kinetic barrier separates the states $\alpha$ and $\beta$.
In this sense, the kinetics is glassy. Therefore, dynamics of HJs
are effectively ergodic within each state on $\mathcal{T}_{obs}$,
but $\mathcal{T}_{obs}$ is not long enough to ensure ergodic sampling
of the entire conformational space $-$ a situation that is reminiscent
of ergodicity breaking in supercooled liquids \cite{ThirumalaiPRA89}.

How many ergodic components states, which do not interconvert among
themselves on $\mathcal{T}_{obs}$, are needed to fully account for
the experimental data? In order to determine the number of states K-means
clustering algorithm was used to partition the conformational space
of HJ into multiple ``ergodic subspaces\textquotedbl{}. This was
achieved by partitioning the stationary distribution of FRET efficiencies,
$p_{s}(E;i)$ = $\lim_{t\rightarrow\mathcal{T}_{obs}}{p(E,t;i)}$,
into distinct states with the requirement that the HJ should ergodically
explore the conformational space within each state. At high Mg$^{2+}$
(50 mM) there are five disjoint states (Fig. \ref{HJFig7}b). The
effective ergodic diffusion constant $D_{E}$ in $E$-space associated
with each state varies greatly from one ergodic subspace to another
(Fig \ref{HJFig7}b).

The HJ gets trapped in one metastable state, which is solely determined
by the initial Mg$^{2+}$ binding \cite{Hyeon12NatChem}. In this
sense, Mg$^{2+}$ plays the role of a random field, which quenches
the conformation of the HJ into one ergodic component. Transition
to another ergodic component can be triggered by using an annealing
protocol in which the Mg$^{2+}$ concentration is first decreased
for a period of time enabling the HJ to explore an entirely different
region of the energy landscape. Subsequent increase of Mg$^{2+}$
concentration results in HJ exploring other ergodic components. The
redistribution of population is clearly shown in Fig.\ref{HJFig7}c
along with the network of connected states. It is indeed surprising
that such a small system exhibit all the key hall marks of slow dynamics
involving multiple ergodic components.

\textit{RecBCD Helicase:} Another example \cite{Liu13Nature} that
vividly illustrates significant molecule-to-molecule variations is
in the function of RecBCD helicase in \textit{E. Coli.}, which is
involved in the repair of breaks in the double stranded DNA (dsDNA)
in an ATP-dependent manner. Here again single molecule experiments
showed that there are dramatic variations in the unwinding speed of
dsDNA depending on the molecule even though all the enzymes were prepared
with no heterogeneity in protein composition. The unwinding velocity,
for specified concentration of ATP, can vary greatly as shown in the
top panel of Fig. \ref{RecBCDFig8}. The most likely explanation is
that the functional landscape is highly heterogeneous with multiple
states each with its own unwinding velocity. This possibility, reminiscent
of the phase space partitioning into ergodic subspaces in Holliday
Junction, was demonstrated using an ingenious set of experiments.
The authors \cite{Liu13Nature} examined the possibility that upon
initially binding Mg$^{2+}$-ATP the enzyme is pinned to one the accessible
states in the functional landscape. In an initial experiment, they
measured the unwinding velocity by incubating the enzyme in the presence
of the ligand, and discovered that RecBCD processively unwinds a large
portion of DNA at a speed that is \textquotedbl{}set\textquotedbl{}
by the initial state. Subsequently, they moved the enzyme to a chamber
without the ligand to stop unwinding for a period of time of about
20 sec. After the period of inactivity, the complex was supplied with
ATP to resume function. Remarkably, the unwinding velocity of the
same molecule changed drastically before and after being depleted
of Mg$^{2+}$-ATP, as shown in the bottom panel of Fig. \ref{RecBCDFig8}.
From the perspective of multiple functional states used to understand
the dynamics of Holliday junction, we can draw three generic lessons
for heterogeneity of RecBCD helicase: (1) The whole space of conformations
partitions into distinct subspaces. The observations that the unwinding
velocity is determined by the dynamics within a single space implies
without changing even after tens of hundreds of base pairs are ruptured,
suggest that the enzyme likely ergodically explores conformations
within a single state. (2) Transitions between distinct states, with
variations in unwinding velocity, can only be achieved by resetting
the ATP concentration, which is reminiscent of Mg$^{2+}$ pulse experiments
used to establish interconversion between distinct states in HJ \cite{Hyeon12NatChem}.
In both cases, ligands act to quench the conformation to a single
substate. Thus, in these systems biological systems heterogeneity
is realized by binding of ligands to the biological molecule. As a
result of pinning the HJ or RecBCD to a single state ergodicity is
effectively broken.

\label{subsec:III.C}

\subsection{Cellular dynamics}

\label{subsec:III.D} Just as is the case in the dynamics of enzymes
and ribozymes discussed above, ensembles averages hide the rich dynamics
associated with cell-to-cell variations. Although the sources of such
variations are hard to pin point except generically as arising from
biochemical noise, as is the case in signaling networks, there is virtually
no question that such variations are manifested in phenotypes \cite{Altschuler10Cell,Alemendro13ARPathol}.
Hence, such stochastic variations are of fundamental importance both
from the perspective of physics as well as biology. In many studies
the behavior of subpopulation of cells are found to be drastically
different from the mean characteristics of the ensemble \cite{Altschuler10Cell},
a situation that is hauntingly similar to DH in glasses. There are
now countless examples of cellular heterogeneity, but here we focus
on one example set in the context of cancer \cite{Alemendro13ARPathol}.
There are apparently profound implications of the observed heterogeneity
including possible variations in the treatment of specific cancers
as it evolves towards metastatic disease - a topic that is far beyond
the scope of the present discussion. We will focus on the similarities
between evolution of cells within a single tumor and particulate glasses.

The variability in cells within tumors differing in the ability to
metastasize and response to drugs were reported long ago \cite{Heppner83Cancer}.
Such variations could arise due to genetic heterogeneity but more
recently it has been appreciated that non-genetic factors including
stochastic variations due to differences in the biochemical reactions,
controlling signaling networks, between cells could also contribute
to cellular heterogeneity (Fig. \ref{CancerFig9}). This has also
been demonstrated most vividly in the differential response of identical
cancer cells to drugs \cite{Alon08Science,Spencer09Nature} or other
therapies. By carefully measuring the expression levels and locations
of a large number of proteins upon treatment of cancer cells with
a drug it was shown that there are dramatic variations in the dynamics
of certain subset of proteins between cells, resulting in the heterogeneous
response. There are substantial variations in the internal stochastic
fluctuations within cells, which manifest themselves as differences
between cells in their response to a cancer drug. In terms of glass
concepts this implies that the various cells can be partitioned (depending
on the the dynamics of individual cells as depicted in Fig. 2A in
\cite{Alon08Science}) into distinct states with distinct dynamics
as shown by huge variations in YFP intensities among different cells.
The similarities to time averaged variations in FRET efficiency between
molecule-to-molecule in Holliday Junction is breath taking.

\section{Glass transition concepts and the RFOT in condensed matter physics}

Typical glassy behavior such as long relaxation times, memory of history,
and physical aging are often observed in the electronic and conductance
properties of low temperature condensed matter systems \cite{Pollack_Ortuno_1985,Ovadyaha_2006}.
There is an enormous amount of experimental and theoretical work on
the glassy behavior in disordered insulators and Coulomb glasses \cite{Pollack_et_al_2013,Amir_et_al_2011}.
More recently it has been appreciated that glassy behavior also occurs
in disordered metallic systems, or electron liquids. For example, below we discuss some aging experiments in an  \textit{metallic} $2D$ MOSFET system. In this system as well as others, see for example the transport properties in the metallic ferromagnet $Sr_{1-x}La_{x}RuO_{3}$ \cite{Kawasaki_et_al_2014}, typical liquid-like glassy behavior is observed.

It is physically
very plausible that a strongly correlated disordered electron liquid
should have many things in common with classical liquids exhibiting
a SGT. First, at least within the RFOT theory of the SGT, the absence or presence of quenched disorder is not important. Second, they are both strongly correlated, frustrated fluids, with identical
spatial symmetries. The frustration in general leads to a rugged energy landscape where concepts such as the Kauzmann transition can play a role. 
In this Section we discuss some connections between the disordered and interacting electron problem and the SGT problem.

We then discuss some theoretical and experimental aspects of super
solids and their connection to what has been referred to as a super
glass \cite{Kim_Chan_2004,Reppy_2010,Ray_Hallock_2009}. Interestingly, the ground state of a interacting Bose system has been related to the Boltzmann measure of a classical hard sphere fluid where RFOT is directly applicable.

Although we focus here on low temperature or quantum condensed matter
systems, there are also very interesting classical or higher temperature
glassy condensed matter systems. For example, recent experiments \cite{Sato_et_al_2014}
in charged cluster glasses have shown a remarkable similarity between
these systems and viscous liquids. Interestingly, as in RFOT there
seems to be an intrinsic relation between dynamics and structure.
Related theoretical work based on RFOT ideas is in \cite{Schmalian_Wolynes_2000}.

\label{sec:IV}

\subsection{Aging in quantum glassy systems }

In general, if $S(t)$ is an observable, or an operator whose quantum
average is an observable, at time $t$ and $h(t)$ is a field conjugate
to $S(t)$ than the correlation function $C$ and the response function
$R$, 
\begin{equation}
C(t,t')=<S(t)S(t')>
\end{equation}
 
\begin{equation}
R(t,t')=\frac{\partial<S(t)>}{\partial h(t')}
\end{equation}
 are related by a fluctuation-dissipation theorem. In addition, in
equilibrium they are functions only of the time difference, $\tau=t-t'$.
In a glassy system, the relaxation is often so slow that on an experimental
time scales neither of these features hold. Let us define $t=\tau+t_{w}$
and $t'=t_{w}$. The time $t_{w}$ is called the aging time, and it
physically represents the duration for which the system was perturbed
before allowing it to relax back to an original equilibrium state.
In non-glassy systems time correlation and response functions do not
depend on $t_{w}$. In all glassy systems, on the other hand, this
history or aging dependence is ubiquitous \cite{Bouchaud_et_al_1997,Struik_1977}.

Next one imagines that $C(\tau+t_{w},t_{w})$ and $R(\tau+t_{w},t_{w})$
consist of two parts: A stationary (ST) part that depends only on
$\tau$ as in non-glassy systems, and an aging (AG) part that depends
on the aging time. For example, we write 
\begin{equation}
R(\tau+t_{w},t_{w})=\mathsf{\mathit{\mathfrak{\mathcal{\mathrm{\mathit{R_{ST}}}}(\tau)+\mathrm{\mathit{R}_{AG}(\tau+\mathit{t_{w}},\mathit{t_{w}})}}}}
\end{equation}
 In general, the precise dependence on the aging time is complicated.
However, deep in the glassy phase there does appear to be a simple
$\tau/t_{w}$ scaling. That is, 
\begin{equation}
R_{AG}(\tau+t_{w},t_{w})\approx F(\tau/t_{w}).
\end{equation}

In Fig.(\ref{ThieryFig10}) \cite{Grenet_et_al_2010} we show the
low temperature ($T=4K)$ conductance, $G(V)$, of insulating granular
aluminum thin films. A ``three-step protocol'' has been used in
these experiments. After the sample is cooled with a gate voltage
$V_{g}=V_{g1}$, a dip forms in $G(V)$ during a time $t_{w1}$, centered
on the voltage $V_{g1}$. The gate voltage is then increased to $V_{g2}$
for a time $t_{w2}$, and a new dip forms while the first one vanishes.
The gate voltage is then changed to $V_{g3}$. The changing of the
dip, $\Delta G_{2}$, at $V_{g2}$ is the measured quantity. $\Delta G_{2}$
can be interpreted as the aging part of the conductance. The important
thing to note is that $\Delta G_{2}$ does depend on the aging times
$t_{w_{1}}$ and $t_{w_{2}}$ and that there is $t/t_{w_{2}}$ scaling.
Again we emphasize that this simple aging phenomena is observed in
numerous classical and quantum systems.

More complicated, or different, aging behavior is observed in disordered
and strongly correlated metallic states. In fact, the change in the
aging behavior from exotic to simple, has been related to the so-called
metal-insulator transition in a two-dimensional electron system in
MOSFETS that were fabricated on the (100) surface of Si. In this system
the crucial quantity is the surface electron density, $n_{s}$. As
$n_{s}$ increases screening improves so that correlation effects
become weaker, and at the same time the disorder is at least partly
weakened since it is in part due to oxide charge scattering which
is also better screened. The first crucial observations was that with
decreasing density there is an apparent metal-insulator transition
at $n_{s}=n_{c}$. Subsequent experiments \cite{Bogdanovich_Popovic_2002}
on the metallic side showed an onset of glassy behavior at $n_{s}=n_{g}$
with $n_{g}>n_{c}$. This second observation showed there was an enormous
increase in the low frequency noise for $n_{s}<n_{g}$ , suggesting
a sudden and dramatic slowing down of the electron dynamics. Later
aging experiments were performed on the same system \cite{Jarosznki_et_al_2007}
.

The system was cooled to either $T=0.5K$ or $T=1.0K$ and an equilibrium
conductivity $\sigma_{o}(n_{s},T)$ was obtained with a gate voltage
$V_{o}$. The gate voltage was then rapidly changed to a different
value $V_{1}$, where it is kept for a time $t_{w}$. The voltage
is then changed back to $V_{o}$, and the slowly evolving $\sigma(t,n_{s},T)$
was measured. The results for $T=1K$ are shown in Fig. (\ref{AgingFig11}).
In the insulating phase, $n_{s}<n_{c}\simeq4.5\pm0.4\times10^{11}cm^{-2}$,
the systems exhibits simple aging with $t/t_{w}$ scaling. In the
metallic glassy region, $n_{c}<n_{s}<n_{g}\simeq7.5\pm0.3\times10^{11}cm^{-2}$,
there is aging with $\mathit{exotic}$ scaling \cite{Kurchan_2002}.
That is, there is apparent $t/t_{w}^{\mu}$ scaling with $\mu$ an
increasing function of $n_{s}$ varying from $\mu=1$, simple aging,
at $n_{s}=n_{c}$ to $\mu\approx3.5$ at $n_{s}=n_{g}$. This scaling
with $\mu>1$ is called super aging. Additional experiments probing DH in these systems would be most interesting.

Super or hyper aging behavior has also been observed in glassy liquids \cite{Leheny_Nagel_1998} and colloidal systems (nano-clay suspensions) \cite{Shanin_Joshi_2012}. It also occurs in various random magnets
and random field-like systems \cite{Bouchaud_et_al_1997,Paul_et_al_2007,Alberici_Kious_et_al_1998}. The connection
between random field problems and RFOT is discussed in Section II.C.

\subsection{Disordered and interacting electrons: Connections with random field
problems and the glass transition problem}

In Section II.C we considered the connection between random field
magnetic problems and the SGT\cite{Biroli_et_al_2013,Franz_et_al_2012,Franz_et_al_2013}.
Here we discuss a connection between random field problems and the
disordered and interacting electron problem \cite{Belitz_Kirkpatrick_1995a,Belitz_Kirkpatrick_1995b,Kirkpatrick_Belitz_1995}.
Physically, since quenched disorder and electron-electron interactions
in general frustrate one another, glassy behavior is anticipated.

\label{subsec:IV.A} Technically, the glassy nature of the interacting
and disordered electron problem is also expected on general grounds.
To see this we start with a schematic action for the problem: 
\begin{equation}
S[\bar{\psi},\psi]=S_{o}+S_{e-e}
\end{equation}
 where $S_{o}$ is the noninteracting, disordered action, 
\begin{equation}
S_{o}=-\sum_{\sigma}\int dx\bar{\psi}_{\sigma}(x)\left[\frac{\partial}{\partial\tau}-\frac{1}{2m}\nabla^{2}-\mu+u(\mathbf{x)}\right]\psi_{\sigma}(x)\label{eq:NI electronic action}
\end{equation}
 and $S_{e-e}$ is the electron-electron interaction term: 
\begin{equation}
S_{e-e}=\frac{\Gamma}{2}\int dx\bar{\psi}_{\sigma_{1}}(x)\bar{\psi}_{\sigma_{2}}(x)\psi_{\sigma_{2}}(x)\psi_{\sigma_{1}}(x)
\end{equation}
 Here $[\bar{\psi},\psi]$ are fermion Grassmann fields, $x=(\mathbf{x},\tau)$
with $\tau$ denoting imaginary time, $\int dx\equiv\int d\mathbf{x}\int_{0}^{1/T}d\tau$,
$m$ is the electron mass, $\mu$ is the chemical potential, $\sigma$
is a spin label, and for simplicity we have assumed an instantaneous
point-like electron-electron interaction with strength $\Gamma$.
$u(\mathbf{x)}$ is a random potential which represents the effects
of disorder. We assume $u$ to be $\delta$-correlated, and obeys
a Gaussian distribution with second moment 
\begin{equation}
\{u(\mathbf{x)}u(\mathbf{y)\}}=\frac{1}{2\pi N_{F}\tau_{el}}\delta(\mathbf{x}-\mathbf{y})
\end{equation}
 where the braces denote the disorder average. Here$N_{F}$ is the
(bare) density of states per spin at the Fermi surface and $\tau_{el}$
is the elastic mean free time.

Theories \cite{Belitz_Kirkpatrick_1994,Belitz_Kirkpatrick_1997} for
the MIT around its lower critical dimension indicate that the natural
order parameter (OP) for the MIT is the single-particle density of
states (DOS) at the Fermi surface, $N$. In terms of the Grassmann
variables this quantity is $N=ImN(i\omega_{n}\rightarrow0+i0)$ ,
with 
\begin{equation}
N(i\omega_{n})=N_{n}=-\frac{1}{2\pi N_{F}}\sum_{\sigma}<\bar{\psi}_{\sigma,n}(\mathbf{x)\psi_{\sigma,\mathrm{n}}(x)}>\label{eq:MIT order parameter}
\end{equation}
 where we have normalized the DOS by $2N_{F}$. Equations (\ref{eq:NI electronic action})
and (\ref{eq:MIT order parameter}) suggest that the OP for the MIT
couples directly to the random potential $u$, and that this random
field (RF) term is structurally identical to the one that appears
in magnetic RF terms. Notice that this term is present in both interacting
and noninteracting disordered electron problems, but in the interacting
case there is an additional physical feature: The interaction term
will in general favor a local electron arrangement that is different
from the one favored by the random potential. This type of frustration
is generally sufficient to lead to glassy behavior.

More formally, the theory using the replica trick to handle the disorder
dependence and the replicated order parameter is (the spin dependence
is suppressed for convenience) 
\begin{equation}
Q_{nm}^{\alpha\beta}(\mathbf{x})=\frac{1}{2}\bar{[\psi}_{n}^{\alpha}(\mathbf{x})\psi_{m}^{\beta}(\mathbf{x})+\bar{\psi}_{m}^{\beta}(\mathbf{x})\psi_{n}^{\alpha}(\mathbf{x})]
\end{equation}
 with 
\begin{equation}
<Q_{nm}^{\alpha\beta}(\mathbf{x})>=\delta_{\alpha\beta}\delta_{nm}N_{n}
\end{equation}
 where now the angular brackets denote both a statistical mechanics
average as well as a disorder average. The random field structure
becomes apparent by transforming the field theory that is originally
in terms of electron operators, to one in terms of the order parameter
$Q^{\alpha\beta}$. Expanding that theory in deviation of $Q$ from
it's average value yields, 
\begin{equation}
Q_{nm}^{\alpha\beta}(\mathbf{x})=<Q_{nm}^{\alpha\beta}(\mathbf{x})>+\phi_{nm}^{\alpha\beta}(\mathbf{x})
\end{equation}
 The resulting theory has an expansion in powers of $\phi$ of the
form, 
\begin{equation}
S=S_{2}+S_{3}+S_{4}+\cdots
\end{equation}
 with $S_{j}\sim\phi^{j}$. Explicitly, 
\[
S_{2}=\int d\mathbf{x}tr\left[\phi(\mathbf{x})\left(-\nabla^{2}+m\right)\phi(\mathbf{x})\right]
\]
 
\begin{equation}
+\frac{\Delta}{2}\int d\mathbf{x}\sum_{i=+,-}(tr_{i}\phi(\mathbf{x}))^{2}
\end{equation}
 with $m$ a mass-like term that is zero at zero frequency and at
the metal-insulator transition (MIT) point where $N(0)$ is vanishes.

At Gaussian order the two point propagator for this theory in the
replica limit is \cite{Belitz_Kirkpatrick_1995a,Kirkpatrick_Belitz_1995},
\begin{equation}
<\phi_{12}(\mathbf{k})\phi_{34}(-\mathbf{k})>=\frac{-4\Delta\delta_{12}\delta_{34}\theta(n_{1}n_{3})}{(k^{2}+m_{n_{1}n_{2}})^{2}}+\cdots
\end{equation}
 where other terms in this correlation function involve only a single
propagator and are therefore less singular. Here $1=\alpha_{1},n_{1}$
etc denotes replica and frequency. This correlation function is characteristic
of a random field problem. Note that there are
cubic terms in this theory, just as there are in the structural glass
random field discussion of Section II.C.

So far an $epsilon$-expansion and ordinary and activated scaling theories of this approach to the MIT have been discussed. It is clear that many aspects of the RF strucure of the MIT need to be investigated. For example, is there a smeared dynamical
glass transition quite apart from the MIT just as in the RFOT of the
SGT transition? Is it related to the glassy behavior observed in the 2D MOSFETS that was discussed in Section
V.A (see also, \cite{Muller_et_al_2012})?

\subsection{The metal-insulator transition and many-body localization}

Apart from the 2D MOSFETS discussed in Section V.A, there has been
an enormous amount of experimental work done on metal-insulator transitions
in three-dimensional interacting and disordered electronic systems.
The subject, however, remains controversial. Significant hysteresis
effects are observed in $Ni(S,Se)_{2}$ and if conventional (as opposed
to activated) scaling is assumed then the dynamical scaling exponent
is surprisingly large \cite{Husmann_et_al_1996}. In the well studied
\cite{Rosenbaum_et_al_1994,Strupp_et_al_1994} doped semiconductor
$SiP$ there are large sample to sample variations that are apparent
only at very low temperatures, $T<60mK,$ suggesting equilibration problems
due to very long relaxation times, and, possibly, dynamical heterogeneity
effects. The glassy aspects of this has been discussed in detail elsewhere
\cite{Belitz_Kirkpatrick_1995b}. Related work on glassy features
of MITs is considered in \cite{Dobrosavljevic_et_al_2012} .

More recently, other glassy aspects of the MIT and interacting and
disordered electrons in general have become apparent. Following ideas
of Anderson \cite{Anderson_1958} , Basko et. al \cite{Basko_et_al_2006}
suggested that it is possible for such a system to remain an insulator
and nonergodic even at a non-zero temperature. Effectively, weakly
interacting localized electrons cannot serve as their own heat bath,
and consequently Mott's variable range hopping doesn't occur in the
absence of delocalized phonons. The basic idea is that since the spectrum
of localized electronic states is discrete the interaction between
electrons will not in general have the $exact$ energy difference
to connect localized states and cause transport. This non-ergodic
phase is called the many-body localized state. Basko et. al further
argued that a system will remain an insulator and non-ergodic up to
a critical temperature they denote by $T_{c}$ and at $T>T_{c}$ the
system will become ergodic and a metal. That is, the MIT occurs at
finite temperature and is a sort of glass transition.

This idea has profound consequences not only for transport theory,
but also for the foundations of quantum statistical mechanics. A basic
tenant of statistical mechanics is that in a big system one can consider
a smaller subsystem and the rest of the system acts as a heat bath
for it. This apparently does not hold in a many-body localized phase.

There has been a a large amount of subsequent work on this problem
\cite{Bauer_Nayak_2013,Huse_Oganesyan_2013,Serbyn_et_al_2013}. Bauer
and Nayak theoretically and numerically investigated the entanglement
entropy of excited states for a system of interacting and disordered
one-dimensional spinless fermions. In the ground state the entanglement
entropy $S(L)$ between a region of size $L$ and the rest of the
system satisfies an area law behaving for large $L$ given by, %
\footnote{This is true for gapped systems. For gapless systems such as Fermi
liquids there are logarithmic corrections.%
}, 
\begin{equation}
S(L)=\alpha L^{d-1}+O(L^{d-2})
\end{equation}
 where $\alpha$ is a constant. This is to be contrasted with highly
excited or thermal states which in general satisfy a volume ($\sim L^{d}$)
law. Importantly, Bauer and Nayak gave evidence that for many-body
localized states the area law holds even for excited states as long
as the interaction strength is not too large. In Fig. (\ref{ChetanFig12})
we show numerical results for excited states for a quantity $a(L)$
that is closley related to the entropy. Here $W$ is a measure of
the disorder and $V$ is a fermion interaction strength. The results
indicate that for large disorder and small interactions the excited
states obey an area entropy law and are thus many-body localized,
but that for smaller disorder and larger interactions the entropy
scale like a volume. This in turn is consistent with a finite temperature
MIT and the considerations of \cite{Basko_et_al_2006}. Remarkably
when this transition is approached from the metallic phase the results
of \cite{Bauer_Nayak_2013} suggest there is a sort of Kauzmann or
RFOT transition characterized by a vanishing entropy at a finite temperature.

\subsection{Super glasses}

In a very interesting paper Biroli, Chamon, and Zamponi (BCZ) \cite{Biroli_et_al_2008,Nussinov_2008}
investigated the so-called super glass phase of matter which is simultaneously
a superfluid and a frozen amorphous structure. Such a system can in
principle be characterized experimentally by placing the system in
a container rotating at a small frequency $\omega$. If the system
is a super solid, and if the frequency is not too high, than one would
find the angular momentum of the solid is reduced from its classical
value $I_{cl}\omega$ by a fraction $f_{s}$ which is called the superfluid
fraction.

BCZ employed a mapping between quantum Hamiltonians and classical
Fokker-Planck operators, to a relate the ground state of a model of
interacting Bosons to the Boltzmann measure of a classical hard sphere
system. They further used this connection and known RFOT results for
the glassy dynamics of Brownian hard spheres to work out the properties
of the super glass phase and the quantum phase transition between
the superfluid and super glass phases. In Fig. (\ref{Francesco1Fig2})
we reproduce their phase diagram summarizing the mapping.

An important experimental question is if pure helium can form an amorphous
phase. Simple monodisperse classical hard sphere systems quickly crystallize
and the glassy phase can only be studied if the quenching rate is
very fast. Superficially one expects the same behavior in helium.
Indeed, path integral Monte Carlo simulations \cite{Biroli_et_al_2011}
of distinguishable $He^{4}$ rapidly quenched from the liquid phase
to very much lower temperatures shows that the system crystallizes
very quickly , without any sign of intermediate glassiness. Interestingly,
it has been suggested that the neglected exchange interaction, and
quantum fluctuations in general, can enhance glassiness.

This last point is very significant and can be understood using RFOT
ideas \cite{Foini_et_al_2010,Foini_et_al_2011} {[}see also \cite{Markland_et_al_2011,Zamponi_2011}{]}.
Consider a classical systems just above the ideal glass transition
temperature with a configurational entropy $S_{c}(s)$ that is a function of the
internal entropy $s$ of the various mosaic states or clusters. The
complexity is small, and in general there will be more compact small
entropy states than large $s$ states. Now add a small amount of quantum
fluctuations as measured by a hopping term $\sim J$. This hopping
will not induce transitions into different mosaic states since that
would involve the movement of a large number of particles which would
be unlikely if $J$ is small. Instead the quantum fluctuations will
cause particle rearrangement within a given cluster. Now small cluster
states cannot easily delocalize to lower their kinetic energy. Instead,
adding the quantum fluctuations will favor larger entropy states that
can more easily delocalize and get bigger. Since these states are
less numerous, $J$ has the effect of decreasing the complexity and
can cause an ideal glass transition.

There is some experimental evidence for both super flow and glassiness
in solid Helium at very low temperatures although the subject remains
very controversial. Using a torsional oscillator experiment Rittner
and Reppy \cite{Rittner_Reppy_2007,Reppy_2010} observed a sample
history dependence with large superfluid fractions ($\sim20\%$) measured
in quenched cooled samples that had small macroscopic dimensions,
and saw reductions of the superfluid fraction that depended on how
much the sample had annealed. This result would be consistent with
a non-equilibrium glassy phase that was not stable. Ray and Hallock
\cite{Ray_Hallock_2008,Ray_Hallock_2009} have performed experiments
in which a chemical potential difference is applied across hcp solid
helium at low densities by injecting liquid helium into one side of
the solid. They observed a dc mass flow at temperatures below approximately
$550$mK. They also observed hysteresis effects: Samples thermally
cycled to, or above, $550$mK do not in general support flow when
cooled down again. This memory effects is consistent with glassy-like
behavior. More experiments are needed in these samples to see if flow
is re-established at still lower temperatures. Still other experiments
are needed to unambiguously confirm or otherwise the super glass phase
of solid helium \cite{Kim12PRB, Mi14}.

\section{Summary and discussion}

\begin{figure*}
\begin{centering}
\includegraphics[width=10cm]{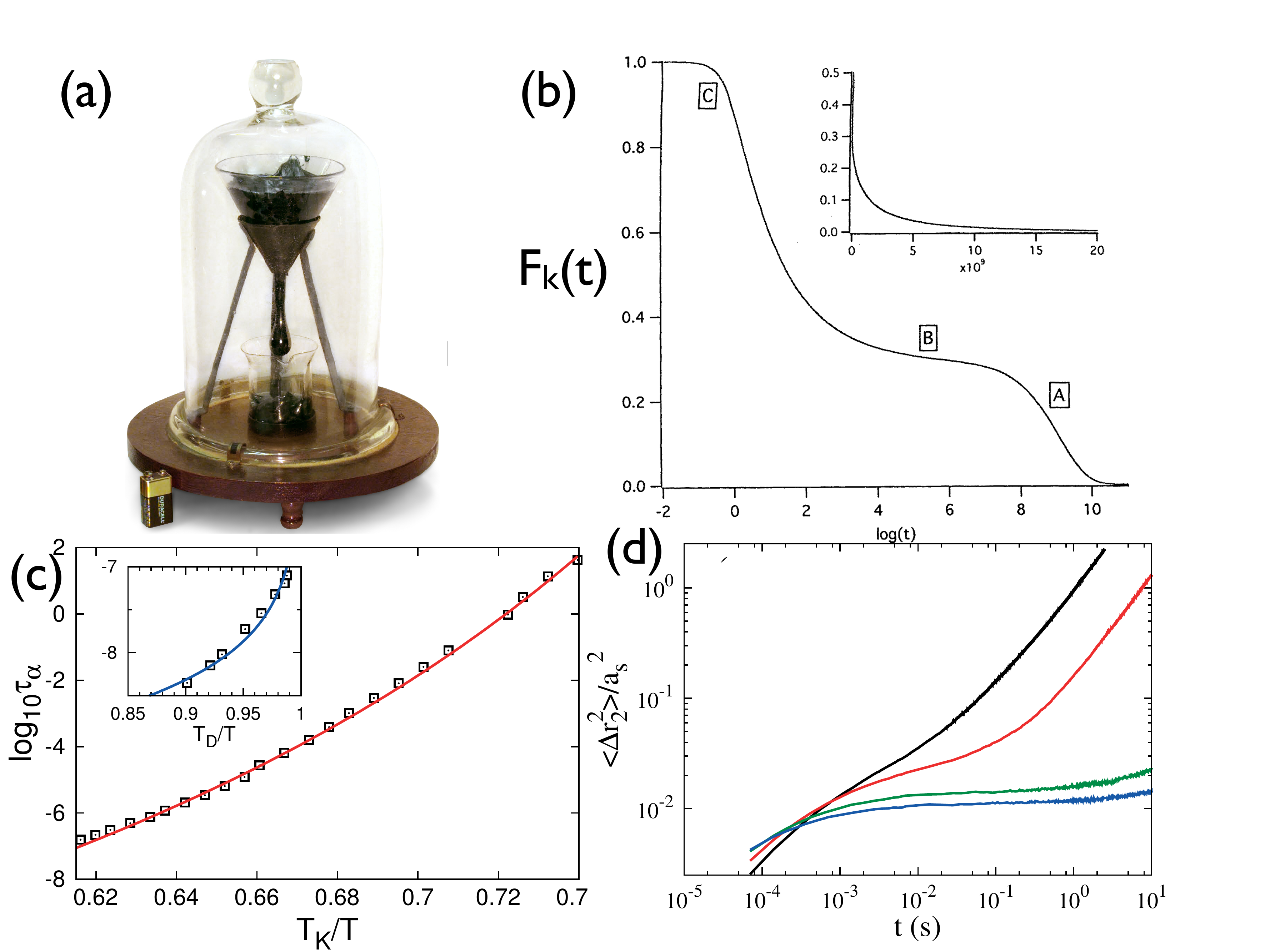} 
\par\end{centering}

\global\long\def\baselinestretch{2}
 \caption{General characteristic of glassy systems. (a) Dramatic illustration
of flow of highly viscous material bitumen in an hour glass. The experiment
was initiated in in 1927 and to date only about nine drops have fallen.
The viscosity of bitumen is about 230 billion times that of water.
(b) Decay of the intermediate scattering function versus log($t$)
is schematically displayed for a slightly undercooled liquid. The
long time $\alpha$-relaxation time is given by regime A and the short
time decay corresponds to regime C. In the intermediate B regime there
is a typically two-step relaxation, which is in accord with the Mode
Coupling Theory. The same plot is displayed in the inset as a function
of $t$, which shows only the long time decay. The figure is adopted
from \cite{Cumminz99JPhysCondMatt}. (c) Dependence of the $\alpha$-relaxation
time, $\tau_{\alpha}$, (regime A in (b)) for salol as a function
of $T_{K}/T$ with the fit given by Eq. \ref{VFT}. The inset shows
$\tau_{\alpha}$ as a function of $T_{D}/T$ where $T_{D}$ is the
dynamical transition temperature. (d) Dependence of the mean square
displacement of a particle as a function of time, $t$, at various
volume fractions for a binary mixture of charge colloidal suspension,
which forms a Wigner glass.}

\label{GlassesGeneralFig1} 
\end{figure*}

\begin{figure*}
\begin{centering}
\includegraphics[width=10cm]{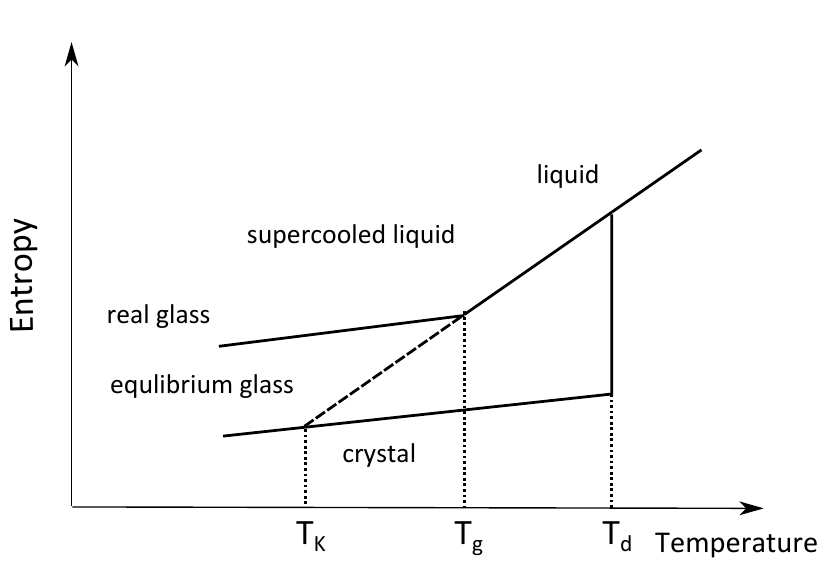} 
\par\end{centering}

\global\long\def\baselinestretch{2}
 \caption{Schematic representation of the configurational entropy change as the temperature of a liquid is reduced. Below the temperature $T_d$, which is an avoided dynamical transition, transport occurs by crossing free energy barriers. At $T < T_g$, the glass temperature, the supercooled liquid falls out of equilibrium. However, if the entropy of the supercooled liquid is extrapolated (dashed line) it would equal the value of the crystal at $T_K$, the Kauzmann temperature.} 

\label{Kauzmann} 
\end{figure*}

\begin{figure*}
\begin{centering}
\includegraphics[width=10cm]{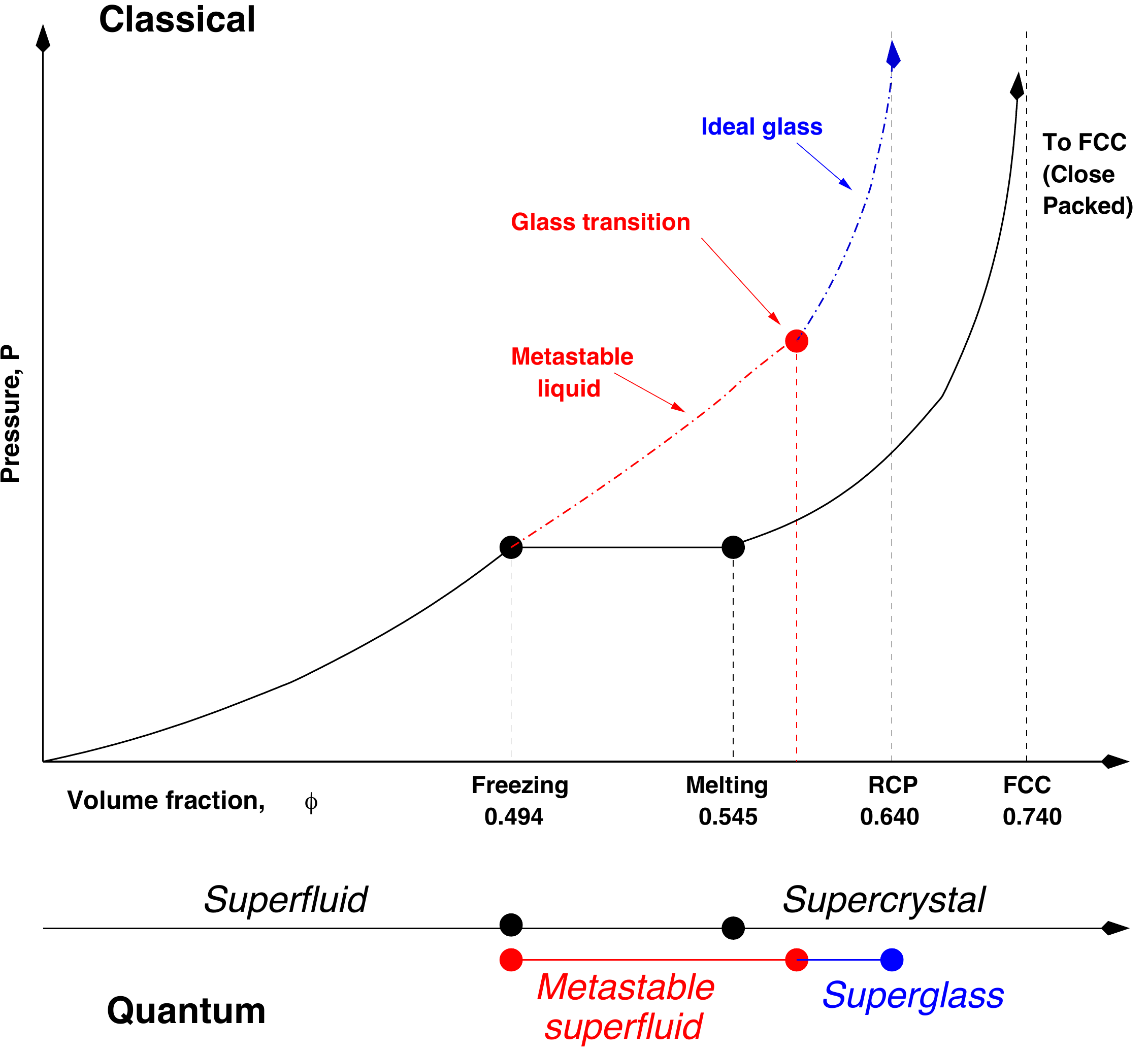} 
\par\end{centering}

\global\long\def\baselinestretch{2}
 \global\long\def\baselinestretch{2}
 \caption{Diagram of states for a classical and quantum hard sphere fluids.
The top panel shows the expected phases as the volume fraction is
increased. Transitions to metastable liquid and glassy phases are
in red. The expected transition to an ideal glassy state, predicted
to occur at $\phi$ close to the random close packing (RCP) is shown
in blue. The lines at the bottom show onset of distinct phases when
quantum effects are taken into account. These are further discussed
in Section V. }

\label{Francesco1Fig2} 
\end{figure*}

\begin{figure*}
\begin{centering}
\includegraphics[width=10cm]{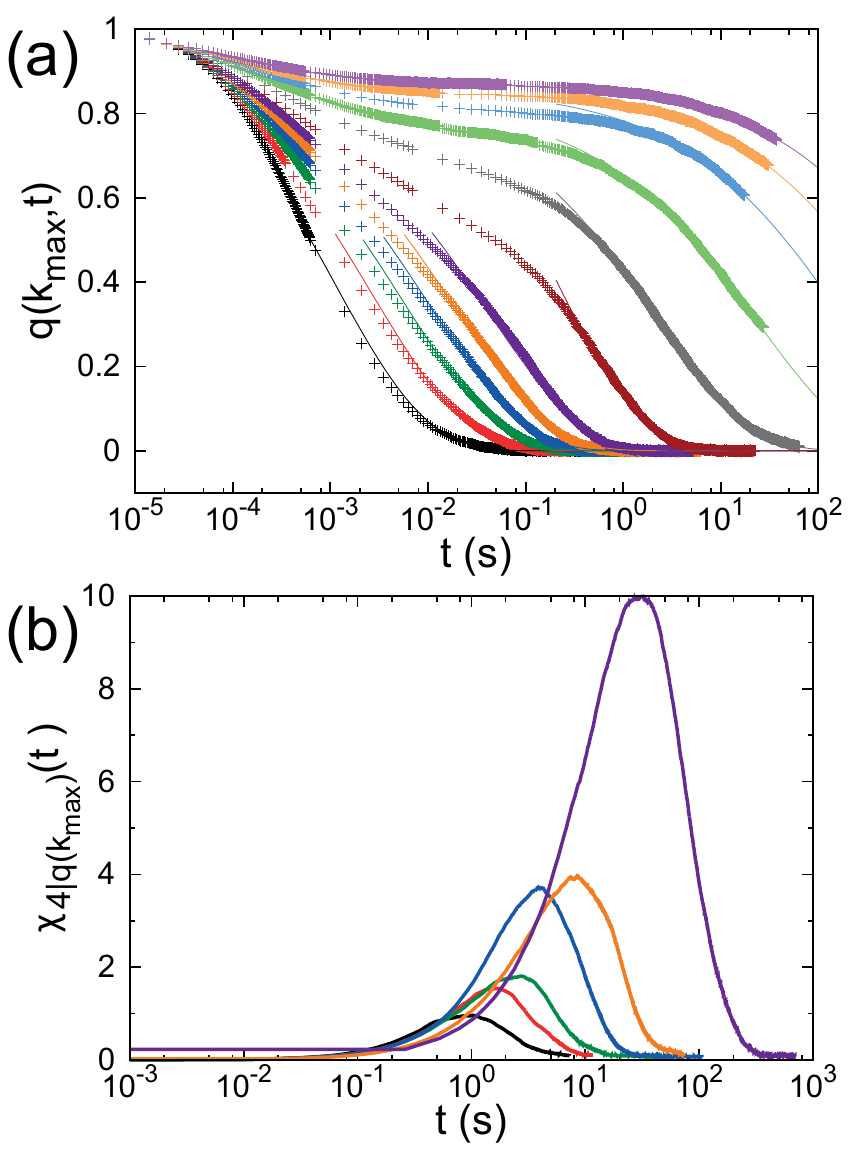} 
\par\end{centering}

\global\long\def\baselinestretch{2}
 \global\long\def\baselinestretch{2}
 \caption{(a) Scattering function for a binary mixture of charged colloidal
suspensions calculated from Brownian dynamics simulations. The mixture
consists of equal number of large and small highly charged spherical
particles. We show $q(k_{max},t)$ for the small particles at $k_{max}$
corresponding to the inverse of the location corresponding to the
first maximum in the pair function. The volume fraction increases
from top to bottom as 0.01, 0.02, 0.03, 0.04, 0.05,
0.06, 0.075, 0.1, 0.125, 0.15, 0.175, and 0.2. The lines are fits to $q(k_{max},t)=exp-(t/\tau_{\alpha})^{\beta}$
with $\phi$-independent $\beta$ = 0.45. (b) Time-dependent changes in the four point
susceptibility, showing fluctuations in $q(k_{max},t)$ for $\phi$ = 0.02, 0.03,
0.04, 0.05, 0.06, and 0.075 from left to right. These figures are adapted from \cite{Kang_Kirkpatrick_Thirumalai_2013}.}

\label{KangFig3} 
\end{figure*}

\begin{figure*}
\begin{centering}
\includegraphics[width=10cm]{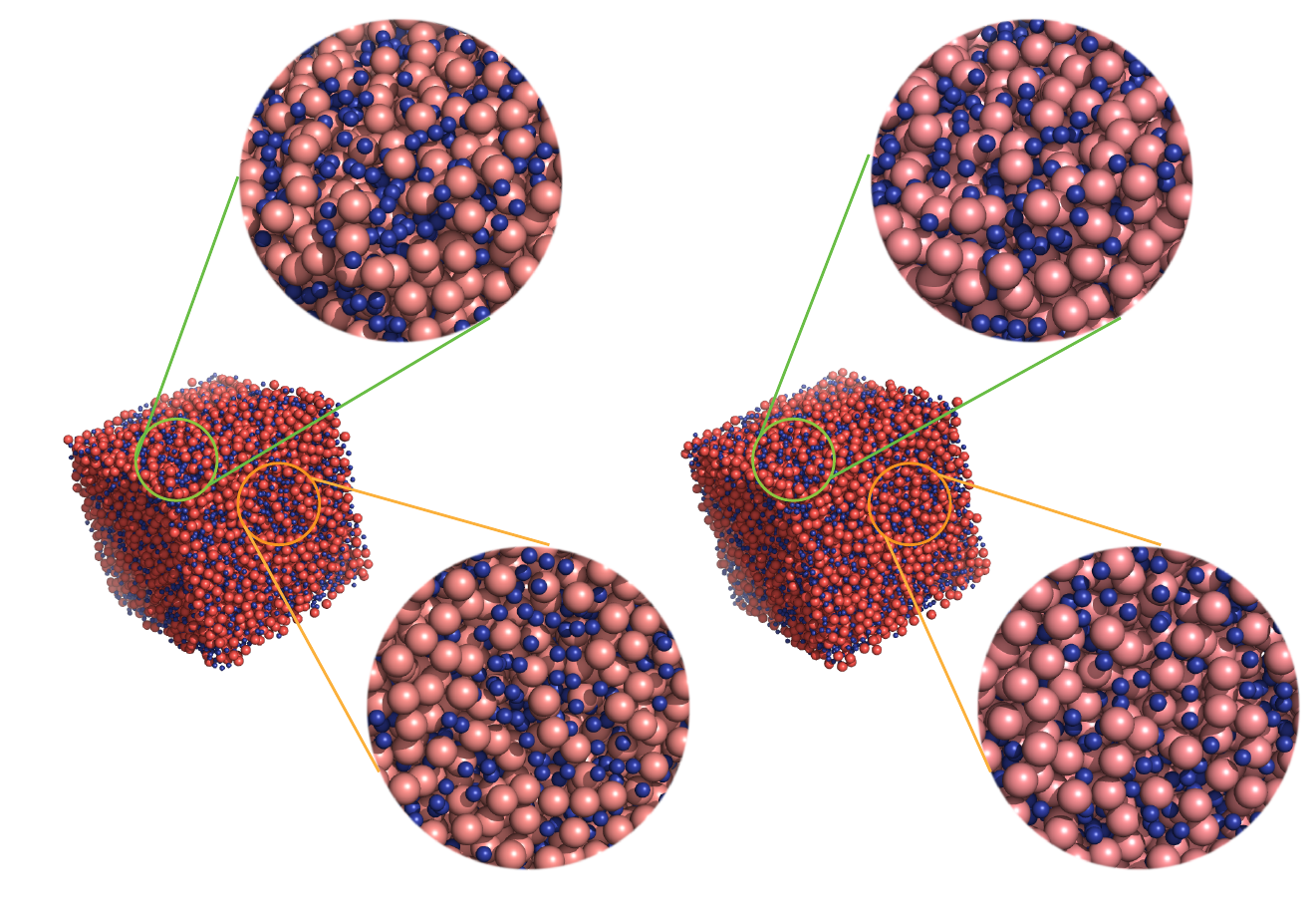} 
\par\end{centering}

\global\long\def\baselinestretch{2}
 \global\long\def\baselinestretch{2}
 \caption{Schematic illustration of dynamic heterogeneity in three dimensional
binary mixture of charged colloidal suspensions. The three figures on the left show a sample prepared at a given time, $t$, for $\phi = 0.10$, which is close to $\phi_d$. The snapshot for two subsamples are shown above and below. The three figures on the right are the same snapshots at a later time $t^{\prime} \approx t + \tau_{\alpha}$. The subsamples on the top are essentially identical where as the ones at the bottom are quite different. Even though there are a large number of particles within each subsample their time evolutions are very different indicating considerable subsample-to-subsample variations. This observation leading to violation of law of large numbers and loss in ergodicity is indicative of dynamical heterogeneity.   These schematic illustrations affirms the mosaic picture of glassy states and shows that only by following the subsamples as a function of time can the extent of dynamic heterogeneity be assessed.
}

\label{DHFig4} 
\end{figure*}

\begin{figure*}
\begin{centering}
\includegraphics[width=7.5in]{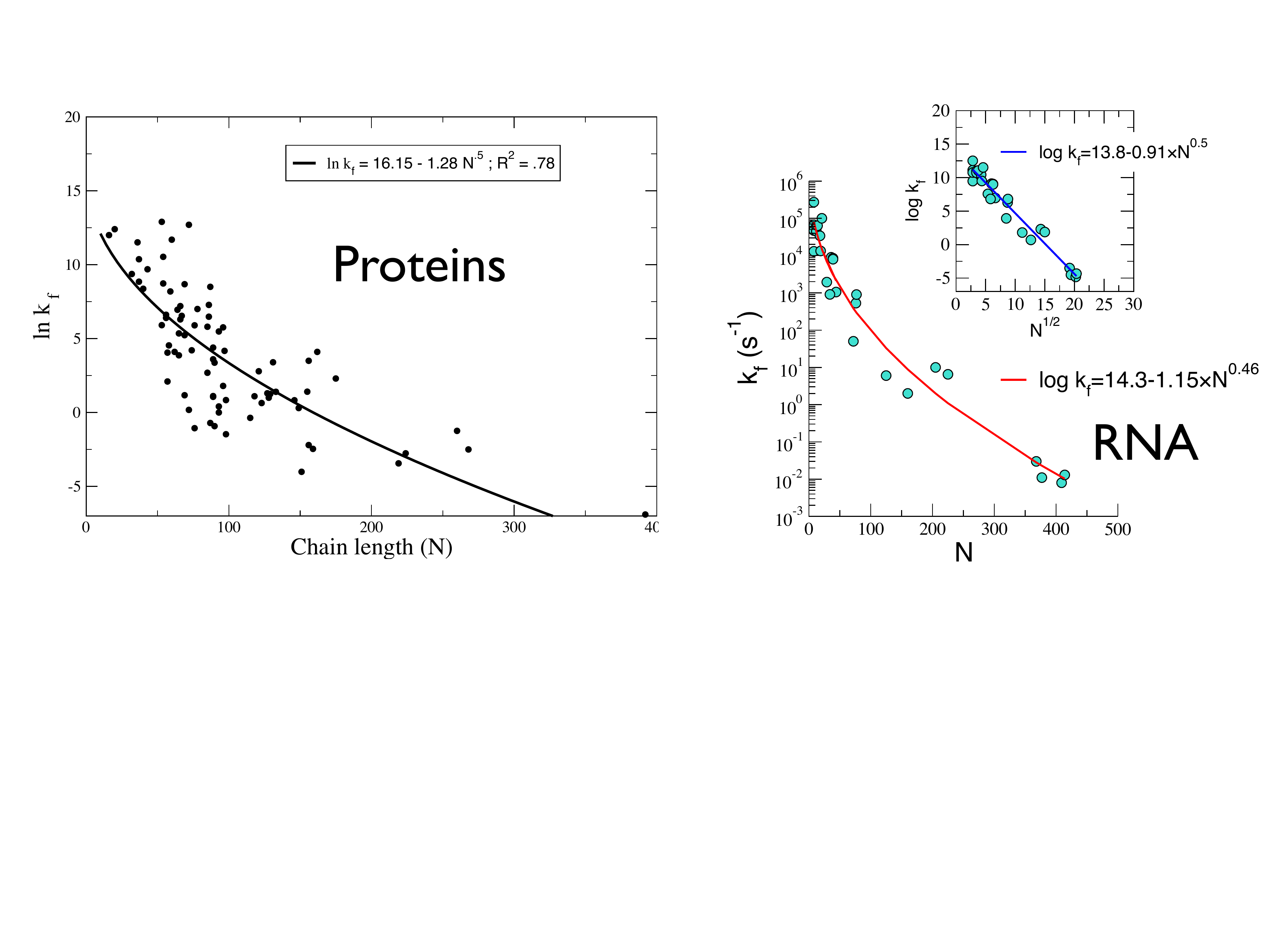} 
\par\end{centering}

\global\long\def\baselinestretch{2}
 \global\long\def\baselinestretch{2}
 \caption{Dependence of the folding rates of proteins (left) and RNA (right)
as a function of length. The fits are based on predictions using $k_{F}\approx e^{\sqrt{(}N)}$,
which follows from activated scaling ideas described in Section IID.}

\label{ScalingFig5} 
\end{figure*}

\begin{figure*}
\begin{centering}
\includegraphics[width=7.5in]{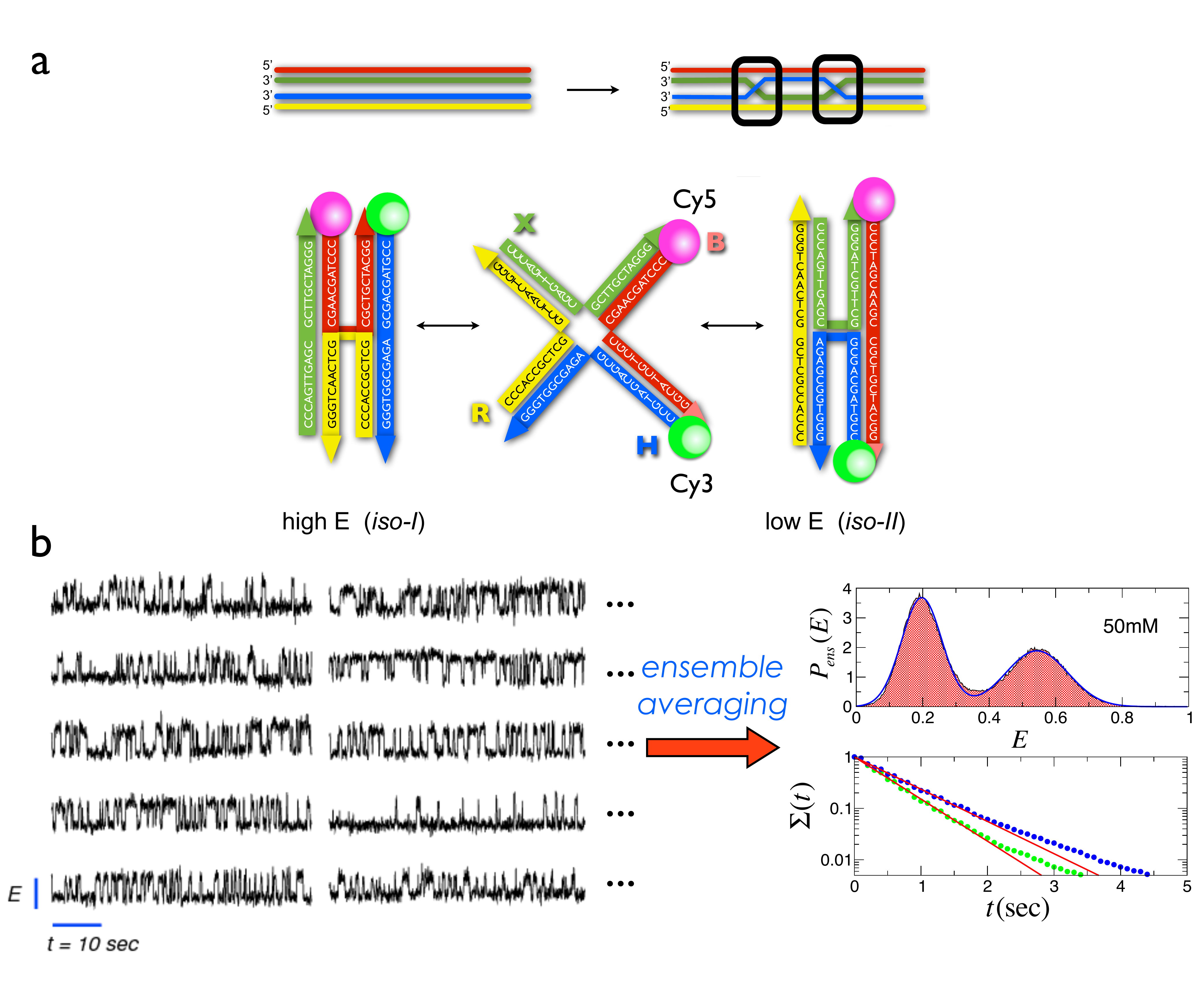} 
\par\end{centering}

\global\long\def\baselinestretch{2}
 \global\long\def\baselinestretch{2}
 \caption{HJ dynamics probed using smFRET experiments. \textbf{a.} Strand exchange
in DNA recombination (top) and the two isoforms of the Holliday Junction
connected by the open square structure (bottom). The Cy5 (magenta)
and Cy3 (green) dyes, attached to the B and H branches in smFRET experiments,
are represented as spheres. \textbf{b.} FRET time traces ($\{E_{i}(t)\}$
with $i=1,2,\ldots,N$ with $N=315$) obtained for individual HJ molecules
at {[}Mg$^{2+}${]} = 50 mM. The ensemble averaged histogram of the
FRET efficiency $E$, i.e., $P_{ens}(E)$, fits to a double-Gaussian
curve (blue line), and the dwell time distribution (bottom panel)
for low (data in green) and high (data in blue) FRET states are approximately
fit to single exponential functions (red lines).}

\label{HJFig6} 
\end{figure*}

\begin{figure*}
\begin{centering}
\includegraphics[width=7.5in]{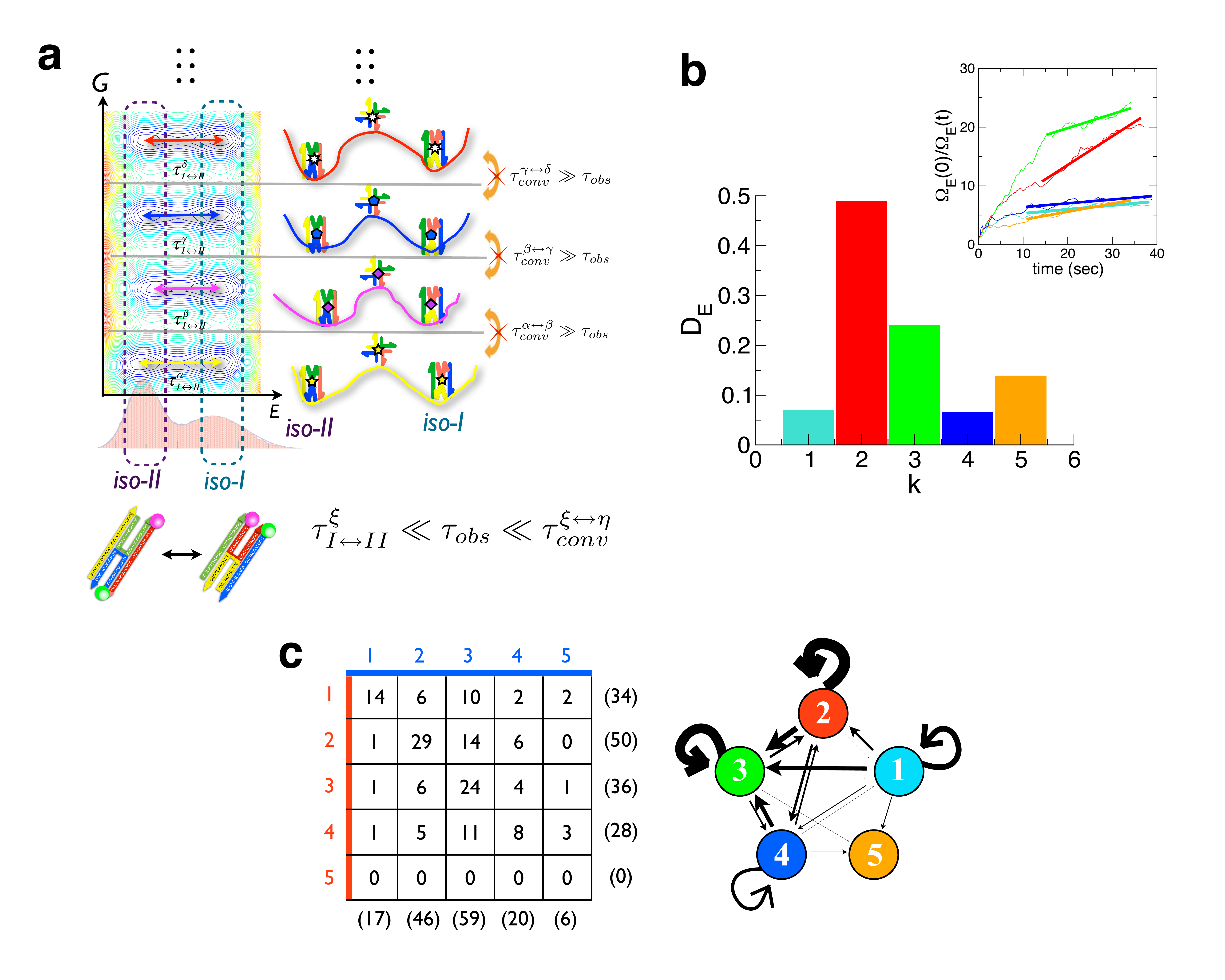} 
\par\end{centering}

\global\long\def\baselinestretch{2}
 \global\long\def\baselinestretch{2}
 \caption{\textbf{a. } Model for the dynamics of HJ constructed based on experiments
and simulations reported in \cite{Hyeon12NatChem}. The free energy
contours for various states are on the left. The isoforms (Fig. \ref{HJFig6}a)
in each state are connected open square form. Ensemble averaged distribution
of the FRET efficiencies, $P_{ens}(E)$, is shown at the bottom. On
the right schematic of the free energy profiles are shown with the
cartoons of HJ structures. The symbols (star, pentagon, $\ldots$)
at the junction emphasize that the junction structure is intact during
the isomerization process. Consequently, $\tau_{\mathrm{I\leftrightarrow II}}^{\xi}$
($\xi=\alpha,\beta\ldots$) $\ll\mathcal{T}_{obs}\ll\tau_{conv}^{\xi\leftrightarrow\eta}$
($\xi,\eta=\alpha,\beta,\gamma,\ldots$ with $\xi\neq\eta$) is established.
\textbf{b.} Five ergodic components are needed to partition the set
of stationary distributions of FRET efficiencies is five. The rates,
$D_{E}$s, of exploration of the conformational space obtained from
the ergodic measure, $\Omega_{E}(t)=\frac{1}{N}\sum_{i=1}^{N}\left(\varepsilon_{i}(t)-\overline{\varepsilon(t)}\right)^{2}$
with $\overline{\varepsilon(t)}\equiv\frac{1}{N}\sum_{i=1}^{N}\varepsilon_{i}(t)$
(shown on top), are different in the distinct ergodic components.
Here, $\varepsilon_{i}(t)$ is the running time average of the FRET
efficiency for molecule $i$, which can be caculated using trajectories
in Fig. \ref{HJFig6}b. \textbf{c.} Evidence for interconversion between
ergodic components by Mg$^{2+}$ reset experiments in 148 molecules.
The indices at the sides of matrix and in the nodes denote the cluster
number $k=1,2\ldots5$. The numbers in the parentheses are the occupation
number in each cluster, which can be obtained by summing up the transition
frequency from one cluster to the other. The diagram on the right
is the kinetic network describing the HJ transition under Mg$^{2+}$
pulse. The widths of the arrows are proportional to the number of
transitions. }

\label{HJFig7} 
\end{figure*}

\begin{figure*}
\begin{centering}
\includegraphics[width=7.5in]{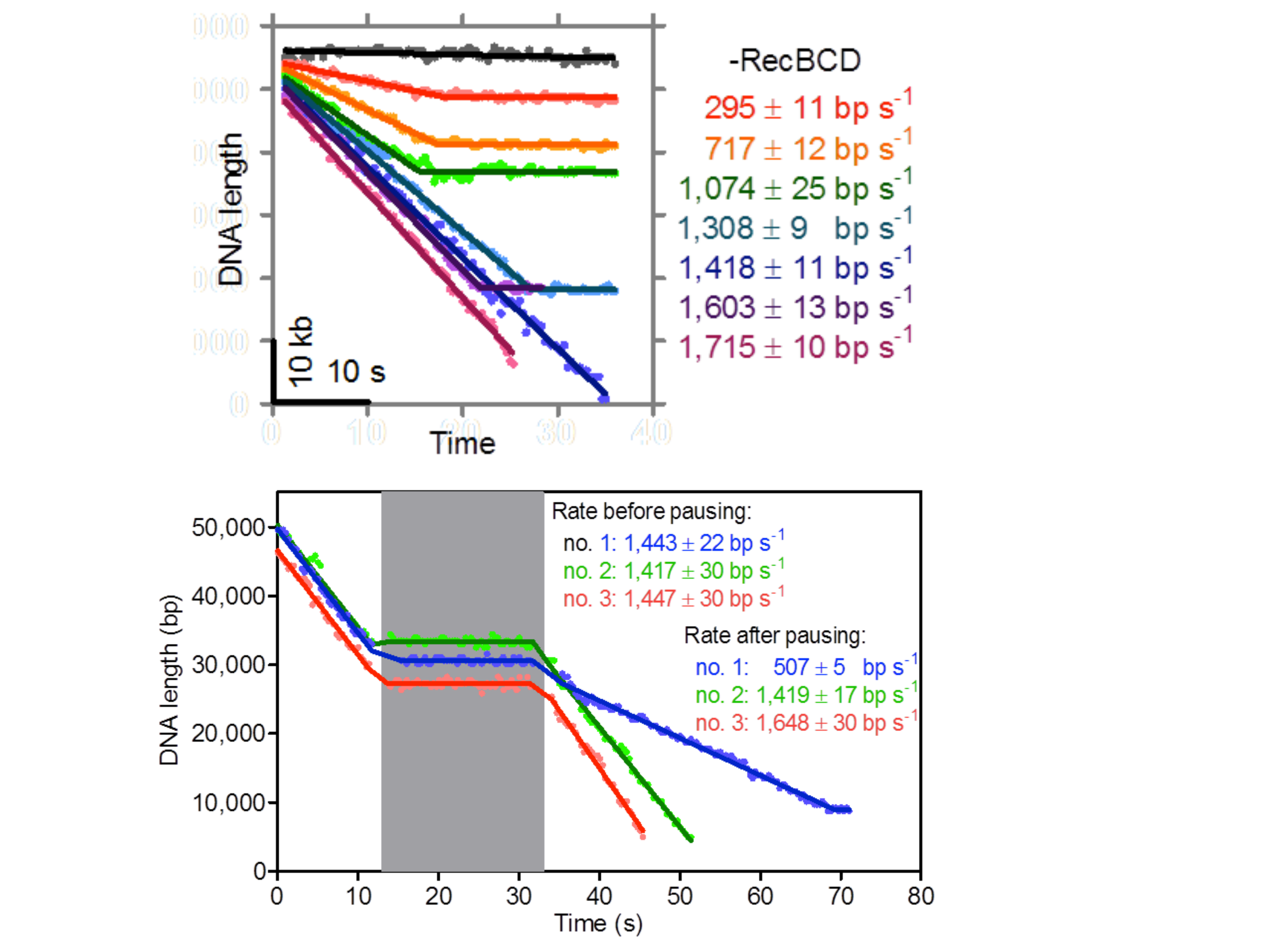} 
\par\end{centering}

\global\long\def\baselinestretch{2}
 \global\long\def\baselinestretch{2}
 \caption{Top panel gives the lenghth of DNA unwound of various RecBCD motors
for various molecules. Black line is for a system without RecBCD.
The unwinding velocities, listed on the right, varies greatly depending
on the molecule. Bottom panel gives the result of reset experiment
in which the motor is depleted of the ligand for a period of time
and reintroduced to resume unwinding. The velocities of the three
motors vary greatly after reset and suggests that the function before
and after reset probe distinct ergodic components.}

\label{RecBCDFig8} 
\end{figure*}

\begin{figure*}
\begin{centering}
\includegraphics[width=7.5in]{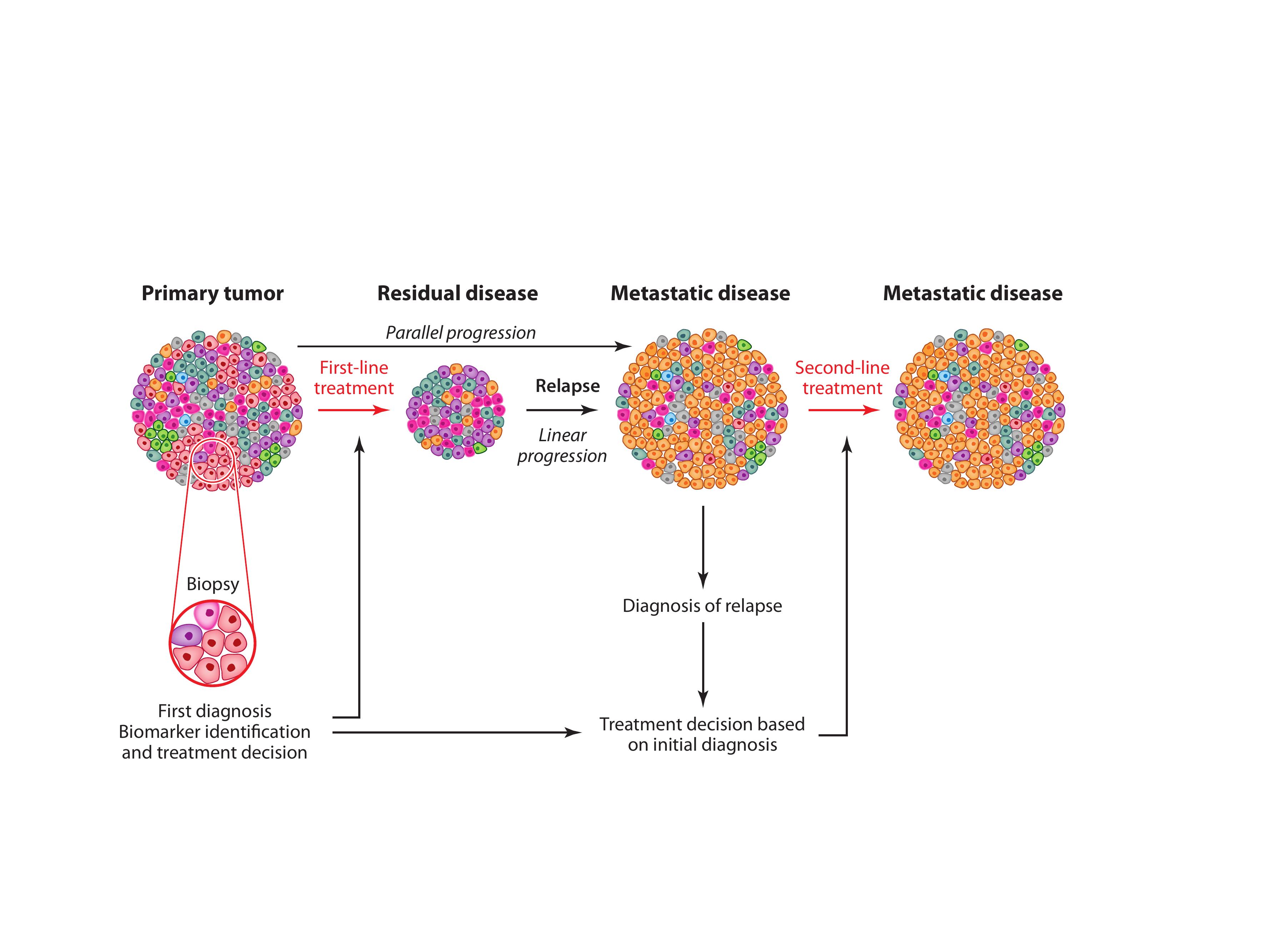} 
\par\end{centering}

\global\long\def\baselinestretch{2}
 \global\long\def\baselinestretch{2}
 \caption{Illustrating intratumor heterogeneity. Typically, cancer diagnosis
is based on sampling a subsample of tumor cells (left hand side of
the figure) baed on biopsy. Because of inherent heterogeneity there
are subsample-to- subsample variations, shown by different colors
on the top left. Treatments based on such biopsies are only successful
in combating the cells in the subsample. Because of stochastic heterogeneity
other clones (shown in yellow) resist the therapy, leading to progression
of the disease. Metastases could develop from clones that survived
the initial therapy. Consequently, treatments based on initial diagnosis
are not efficacious in fighting proliferation at subsequent times,
which is an inherent feature of heterogeneity much like in glasses. }

\label{CancerFig9} 
\end{figure*}

\begin{figure*}
\begin{centering}
\includegraphics[width=7.5in]{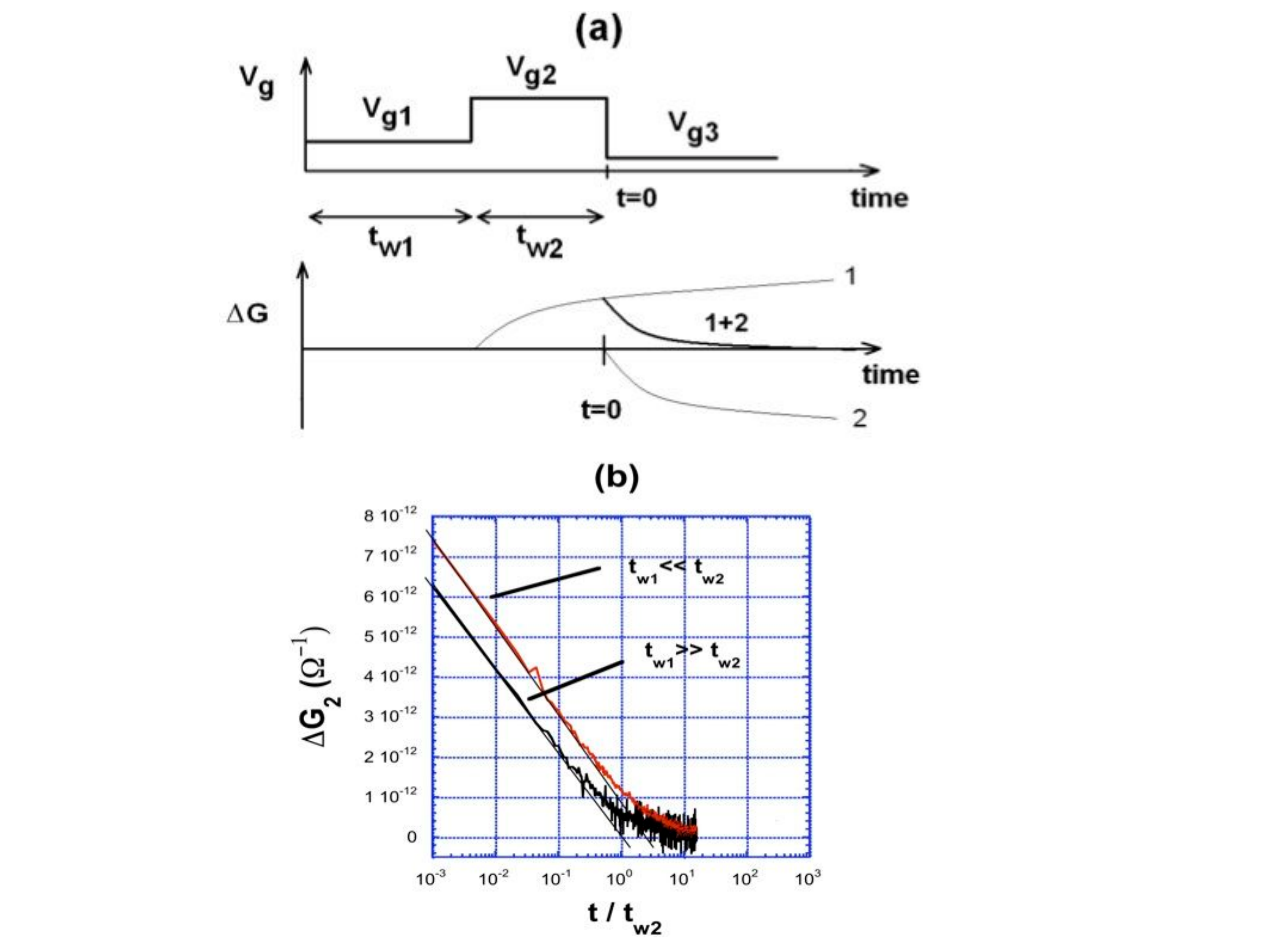} 
\par\end{centering}

\global\long\def\baselinestretch{2}
 \global\long\def\baselinestretch{2}
 \caption{ (a) Experimental protocol as described in text. (b) Gate voltage
dip as a function of time $t>0$ for $t_{w_{1}}\ll t_{w_{2}}$ and
$t_{w_{1}}\gg t_{w_{2}}$.}

\label{ThieryFig10} 
\end{figure*}

\begin{figure*}
\begin{centering}
\includegraphics[width=7.5in]{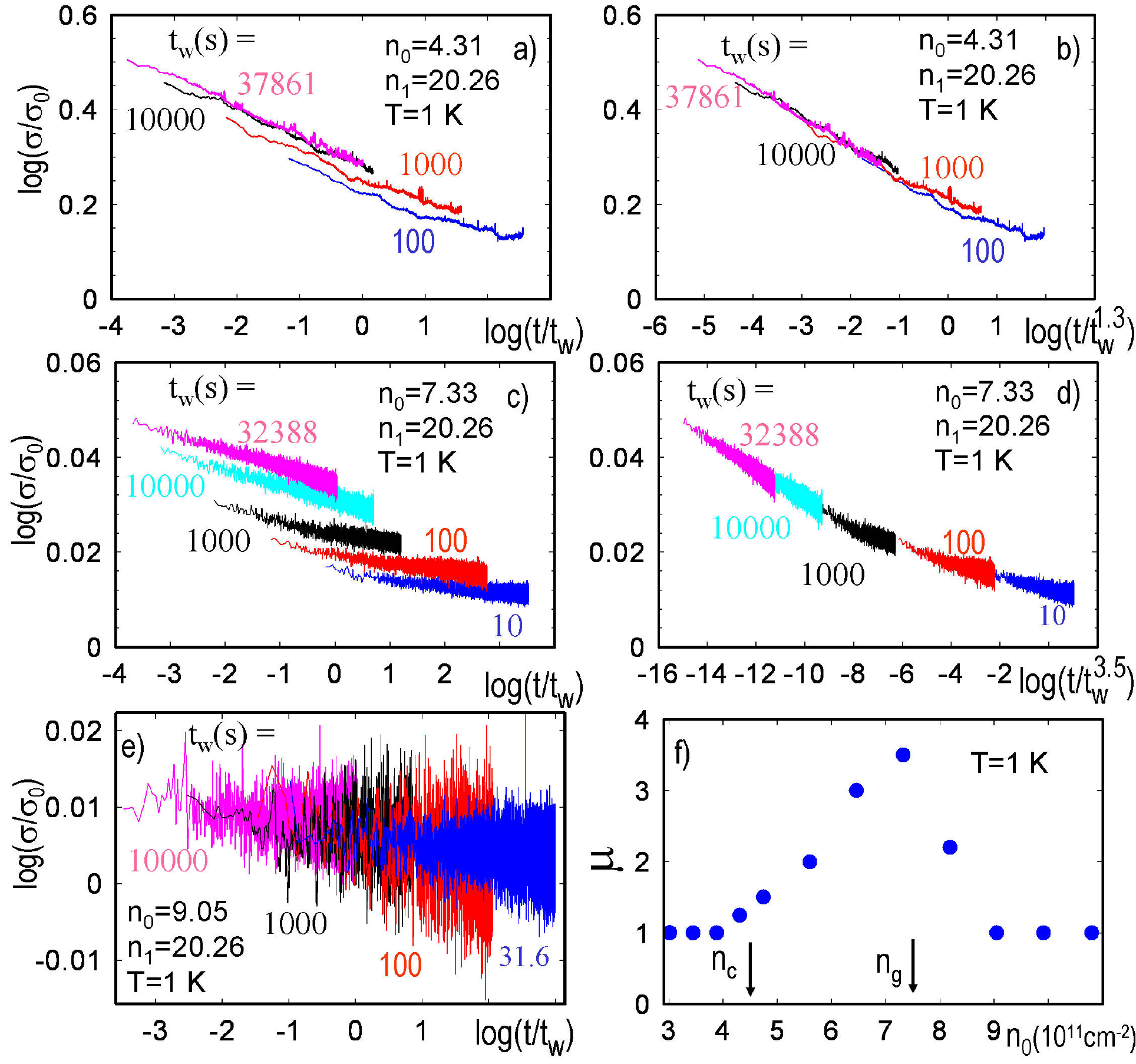} 
\par\end{centering}

\global\long\def\baselinestretch{2}
 \global\long\def\baselinestretch{2}
 \caption{ 2D MOSFET with initial gate voltage $V_{0}$(density $n_{0}$) that
is changed to a voltage $V_{1}$(density $n_{1}$) for a time $t_{w}$
and then changed back to voltage $V_{0}$ at a time $t=0$. $\sigma(t>0,t_{w})$
is measured. (a), (c), (e) Relaxations for different $n_{0}$ at fixed
$n_{1}$, scaled with time $t_{w}$. (b), (d) Scaling with $t_{w}^{\mu}$
improves the collaspe of the data. (f) $\mu$ vrs $n_{0}$ does not
depend on $n_{1}$. }

\label{AgingFig11} 
\end{figure*}

\begin{figure*}
\begin{centering}
\includegraphics[width=7.5in]{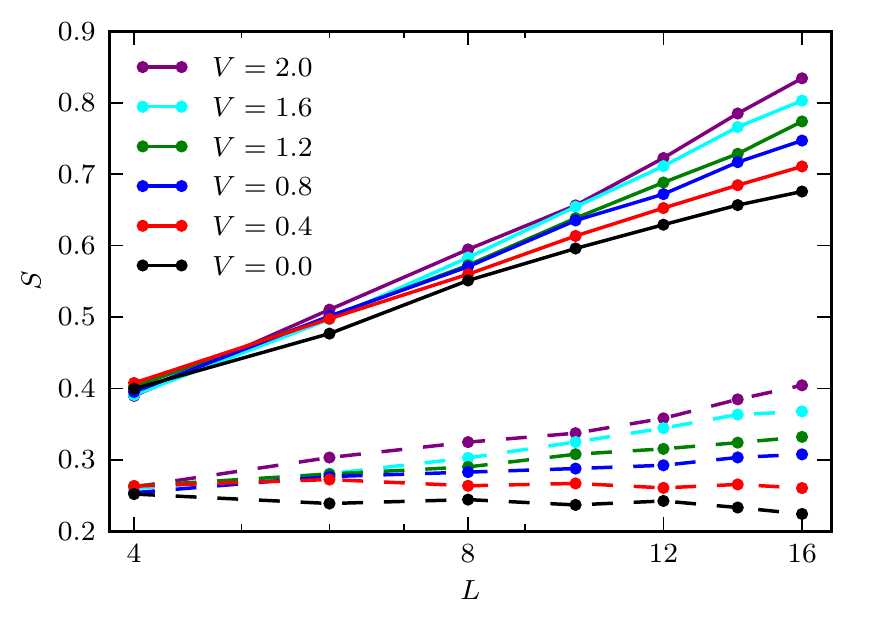} 
\par\end{centering}

\global\long\def\baselinestretch{2}
 \global\long\def\baselinestretch{2}
 \caption{The coefficient $a(L)$ is a measure of the entropy. Solid lines are
$W=8$, dashed lines are $W=6$ ; interaction strength is $V=0,0.4,0.8,1.2,1.6,2.0$.
In a many body localized regime $a(L\rightarrow\infty)$ approaches
a constant, while in a metallic state $a(L\rightarrow\infty)$ grows
linearly with system size. }

\label{ChetanFig12} 
\end{figure*}

All of the themes that we have highlighted in this article, which
can be viewed from the perspective of concepts developed in glass
physics, are active fields of research. It should be emphasized that
our viewpoint is not universally endorsed, and hence there is a spirited
debate on the origins of slow relaxation in glasses. It is unclear
if there is an underlying structural order parameter describing the
stability or dynamics glass forming systems. The search for such order
parameter has been pursued for nearly thirty years, and it has been
asserted that some sort of orientational order may increase upon supercooling.
However, such a conclusion may only be relevant to quasi one component
systems but the generality of this notion for complex glass forming
materials is hardly obvious. In addition, the unambiguous demonstration
of the existence of an the ideal glass transition temperature ($T_{K}$
in the VFT fit) in experiments has been very difficult. For example,
fitting viscosity data for Salol Fig.\ref{GlassesGeneralFig1}a shows
that temperatures at which reliable measurements can be made are far
from $T_{K}$ with $\epsilon\approx$ 0.26 the dimensionless distance
from the transition. It is even more difficult to show $T_{K}\ne0$
in computer simulations although the plausibility of a thermodynamic
transition envisioned in RFOT has been hinted at using random pinning
simulations \cite{Karmakar13PNAS,Kob13PRL}.

Despite these reservations in three key papers \cite{Kurchan_et_al_2012,Kurchan_et_al_2013,Chabonneau_et_al_2013}{[}see
also \cite{Kirkpatrick_Wolynes_1987a}{]} , have studied in detail
the thermodynamics of hard sphere particles in large dimensions ($d=\infty$)
and all of the predictions of RFOT have been exactly demonstrated.
By exploiting the observation that in this system at $d=\infty$ only
the second virial coefficient contributes to the free energy functional
of the system it was shown that the one step replica symmetry breaking
(1RSB) and the two transitions (with the variable being density as
opposed to temperature) scenario, as anticipated in the RFOT theory
\cite{Kirk89JPhysA}, is valid. In addition, they discovered an instability
of the 1RSB at high density resulting in the Gardner transition. It
is generally believed that an inherently mean field description is
reasonable for liquids (except close to gas-liquid critical point),
and hence the large dimensional theory may have wider range of applicability
(see, for example \cite{Kirkpatrick_1986,Marechal_et_al_2012}.

We have barely touched on the potential application of glass transition
concepts in biological problems. One noteworthy example 
 is the folding of chromosomes, which could result in manifestation of metastability and glass-like
behavior due to topological constraints \cite{Hyeon11NatComm}.   In eukaryotic cells chromosomes
fold into globules occupying well-defined regions referred to as chromosome
territories \cite{cremer2001NRG}, thus bringing widely separated
gene-rich regions are brought into close proximity. Folding of chromosomes
apparently occurs without forming knots, which is important for gene
activity, in a polymer containing many mega base pairs. Using constraints
derived from experiments as a guide \cite{lieberman09Science} it
has been argued that the genome is packaged into fractal globules
\cite{grosberg1993EPL} differing qualitatively from equilibrium globules
in which formation would occur with high probability. It is most likely
the case that there are multiple states associated with fractal globules,
which implies that the dynamics of chromosome folding would be glassy.
Although the biological implications are unclear, it is worth exploring
genome folding in various eukaryotic cells to assess if glass like
behavior is exhibited, and to understand if nature utilizes such dynamics
in of the most crucial functions.

There is an enormous amount of glassy phenomena that occur in the so-called
hard condensed matter physics systems. Generally these are quantum
systems at low temperatures. They include Coulomb glasses, disordered
insulators, disordered metals, quantum phase transitions from a superconducting
state to either a disordered insulator or metal, various non-Fermi
liquid systems, quantum Griffith's phase effects, etc. Even in  low temperature ferromagnetic metals there are numerous
manifestations of glassy effects \cite{Kawasaki_et_al_2014}. One of the main problems is that
there is not a common language, let alone a common description, in
these various subfields. It is possible that some of the recent unifying
ideas in classical glassy systems will be relevant in these quantum
systems. For example, in the SGT problem there has been a tremendous
amount of work recently on dynamic heterogeneity. There has been a
fruitful interplay between theory, simulations and experiments. This
concept is also clearly relevant in biological glassy systems, as
illustrated here. Recently, it has been shown that single molecule
pulling experiments on proteins and DNA provide direct evidence for
heterogeneity on the molecular scale \cite{Hyeon14PRL}. In the condensed
matter case this subject has hardly been touched \cite{Nussinov_et_al_2013}.
In understanding the similarities and differences between classical
and quantum glassiness two fundamental differences must be kept in
mind. The first is that quenched disorder is perfectly correlated
along the imaginary time direction and this can have especially profound
implications for quantum phase transitions \cite{Vojta_2006}. If
the disorder is self-generated as in the case of the SGT, it is likely
that similar profound effects will occur. The second is that in general
there are modes that are soft only at $T=0$, and these extra soft
modes \cite{Belitz_Kirkpatrick_2014} will likely play an important
role in the long time glassy dynamics.


\vskip 400pt \acknowledgments {We are grateful to Changbong Hyeon,
Hongsuk Kang, Fracesco Zamponi, Stephen Kowalczykowski, Kingsuk Ghosh,Thierry Grenet,
Dragana Popovic, and Chetan Nayak for providing us figures for reproduction. We thank Hongsuk Kang for useful discussions.
We are grateful to the National Science Foundation for supporting
this work through Grants No. CHE 13-61946 and No. DMR-09-01907.} 
\newpage

 \clearpage{} 

%

\clearpage{} 

\newpage

\clearpage

\newpage{} 

\newpage{} 

\newpage{} 

\newpage{} 

\newpage{} 

\newpage{} 

\newpage{} 

\newpage{} 

\newpage{} 

\newpage{} 

\end{document}